\documentclass[
 aps,
 prb,
 reprint,
 amsmath,
 amssymb,
 superscriptaddress,
floatfix
]{revtex4-2}
\usepackage{graphicx}

\usepackage[utf8]{inputenc}
\usepackage[T1]{fontenc}

\usepackage{siunitx}
\DeclareSIUnit\angstrom{\protect \text {Å}}

\usepackage{hyperref}
 \hypersetup{
     colorlinks   = true,
     citecolor    = blue,
     linkcolor    = blue,
     urlcolor     = blue
}

\PassOptionsToPackage{version=3}{mhchem}
\usepackage{mhchem}

\newcommand{\TN}{T_\text{N}}
\newcommand{\TC}{T_\text{C}}
\newcommand{\EF}{E_\text{F}}

\newcommand{\Tstar}{T^*}
\newcommand{\TFIF}{T_\text{FIF}}

\newcommand{\cax}{\ensuremath{c}~axis}

\newcommand{\muH}{\mu_0H}

\newcommand{\muat}{\mu_\text{at.}}

\newcommand{\EZP}{\ce{EuZn2P2}}
\newcommand{\Euion}{\ce{Eu^{2+}}}

\newcommand{\Hac}{H_\text{ac}}

\newcommand{\mueff}{\mu_\text{eff}}
\newcommand{\NA}{N_\text{A}}
\newcommand{\kB}{k_\text{B}}
\newcommand{\Thp}{\theta_\text{p}}
\newcommand{\muB}{\mu_\text{B}}

\newcommand{\ThD}{\theta_\text{D}}
\newcommand{\ThEi}{\theta_{\text{E}_i}}

\newcommand{\Cmag}{C_{\text{mag}}}
\newcommand{\Cph}{C_{\text{ph}}}
\newcommand{\Smag}{S_{\text{mag}}}
\newcommand{\Tad}{T_{\text{ad}}}

\newcommand{\Hinitial}{H_{\text{i}}}
\newcommand{\Hfinal}{H_{\text{f}}}

\newcommand{\Bhf}{B_{\text{hf}}}

\begin{document}

\title{Ambient and high pressure studies of structural, electronic and magnetic properties of EuZn$_2$P$_2$ single crystal}

\author{Damian Rybicki}
\email[Corresponding author: ]{ryba@agh.edu.pl}
\affiliation{AGH University of Krakow, Faculty of Physics and Applied Computer Science, al. A. Mickiewicza 30, 30-059 Krak\'ow, Poland}

\author{Kamila Komędera}
\affiliation{AGH University of Krakow, Faculty of Physics and Applied Computer Science, al. A. Mickiewicza 30, 30-059 Krak\'ow, Poland}

\author{Janusz Przewoźnik}
\affiliation{AGH University of Krakow, Faculty of Physics and Applied Computer Science, al. A. Mickiewicza 30, 30-059 Krak\'ow, Poland}

\author{Łukasz Gondek}
\affiliation{AGH University of Krakow, Faculty of Physics and Applied Computer Science, al. A. Mickiewicza 30, 30-059 Krak\'ow, Poland}

\author{Czesław Kapusta}
\affiliation{AGH University of Krakow, Faculty of Physics and Applied Computer Science, al. A. Mickiewicza 30, 30-059 Krak\'ow, Poland}

\author{Karolina Podgórska}
\affiliation{AGH University of Krakow, Faculty of Physics and Applied Computer Science, al. A. Mickiewicza 30, 30-059 Krak\'ow, Poland}

\author{Wojciech Tabiś}
\affiliation{AGH University of Krakow, Faculty of Physics and Applied Computer Science, al. A. Mickiewicza 30, 30-059 Krak\'ow, Poland}

\author{Jan Żukrowski}
\affiliation{AGH University of Krakow, Faculty of Physics and Applied Computer Science, al. A. Mickiewicza 30, 30-059 Krak\'ow, Poland}

\author{Lan Maria Tran}
\email[Corresponding author: ]{l.m.tran@intibs.pl}
\affiliation{Institute of Low Temperature and Structure Research, Polish Academy of Sciences, Wroc\l{}aw, 50-422, Poland}

\author{Micha\l{} Babij}
\affiliation{Institute of Low Temperature and Structure Research, Polish Academy of Sciences, Wroc\l{}aw, 50-422, Poland}
\author{Zbigniew Bukowski}
\email[Deceased]{}

\affiliation{Institute of Low Temperature and Structure Research, Polish Academy of Sciences, Wroc\l{}aw, 50-422, Poland}

\author{Ladislav Havela}
\email[Corresponding author: ]{ladislav.havela@matfyz.cuni.cz}
\affiliation{Charles University, Faculty of Mathematics and Physics, Department of Condensed Matter Physics, Ke
Karlovu 3, 121 16 Prague 2, Czech Republic}

\author{Volodymyr Buturlim}
\affiliation{Charles University, Faculty of Mathematics and Physics, Department of Condensed Matter Physics, Ke
Karlovu 3, 121 16 Prague 2, Czech Republic}

\author{Jiri Prchal}
\affiliation{Charles University, Faculty of Mathematics and Physics, Department of Condensed Matter Physics, Ke
Karlovu 3, 121 16 Prague 2, Czech Republic}

\author{Martin Divis}
\affiliation{Charles University, Faculty of Mathematics and Physics, Department of Condensed Matter Physics, Ke
Karlovu 3, 121 16 Prague 2, Czech Republic}

\author{Petr Kral}
\affiliation{Charles University, Faculty of Mathematics and Physics, Department of Condensed Matter Physics, Ke
Karlovu 3, 121 16 Prague 2, Czech Republic}

\author{Ilja Turek}
\affiliation{Charles University, Faculty of Mathematics and Physics, Department of Condensed Matter Physics, Ke
Karlovu 3, 121 16 Prague 2, Czech Republic}

\author{Itzhak Halevy}
\affiliation{Ben Gurion university, Be’er Sheva, Israel}

\author{Jiri Kastil}
\affiliation{Institute of Physics, Academy of Science of the Czech Republic, Prague 8, Czech Republic}

\author{Martin Misek}
\affiliation{Institute of Physics, Academy of Science of the Czech Republic, Prague 8, Czech Republic}

\author{Dominik Legut}
\affiliation{Charles University, Faculty of Mathematics and Physics, Department of Condensed Matter Physics, Ke
Karlovu 3, 121 16 Prague 2, Czech Republic}
\affiliation{IT4Innovations, V\v{S}B - Technical University of Ostrava, 17. listopadu 2172/15, 70800 Ostrava, Czech Republic}

\begin{abstract}
A thorough study of \EZP{} single crystals, which were grown from Sn flux, was performed using both bulk (heat capacity, ac susceptibility, dc magnetization, electrical resistivitivity, magnetoresistance) and microscopic (M\"ossbauer spectroscopy) techniques. 
Electrical resistance and magnetic susceptibility were measured also under high pressure conditions (up to 19 GPa and 9.5 GPa, respectively).  
Further insight into electronic properties and phonons is provided by \textit{ab initio} calculations. 
The results indicate that \EZP{} is an antiferromagnet with strong Eu-Eu exchange coupling of ferromagnetic type within the basal plane and weaker antiferromagnetic interaction along the $c$ axis. The Eu magnetic moments are tilted from the basal plane.  
Hydrostatic pressure strongly affects both magnetic (increase of the N\'{e}el temperature) and electronic (suppression of the band gap and semi metallic behavior) properties, indicating a strong interplay of structure with magnetic and electronic degrees of freedom.
\end{abstract}

\maketitle

\section{Introduction}
Zintl-like phases can be seen as compounds of an electropositive cation, which donates valence electrons to form a framework of covalently bonded poly-atomic anions with a closed valence shell. Because of the nature of the chemical bonding, Zintl phases are located at the borderline between valence compounds, which are typical insulators, and the intermetallic compounds, which are typical metals. Often exhibiting intermediate properties, Zintl phases are typically narrow band gap semiconductors. The concept of Zintl like phases was stretched over the periodic table of elements and many Zintl compounds containing $d$ and $f$ metals are reported in the literature. The situation of a tunable narrow band gap is interesting for practical reasons, as it gives, e.g., prominent magnetocaloric properties~\cite{kauzlarich_zintl_2023}.

A specific situation providing an excellent opportunity of well-defined anisotropy of magnetic and transport properties is offered by the trigonal structure of the CaAl$_2$Si$_2$ type (space group No. 164, \textit{P}-3\textit{m}1), built by alternating cationic (Ca) and anionic (Al-Si) layers~\cite{Zheng1988}. Incorporating magnetic atoms (e.g.  Eu) in such a compound gives an intriguing opportunity to explore interplay of magnetic degrees of freedom and electronic structure. In particular, magnetic moments reorientation (in magnetically ordered state), or disordering (in the paramagnetic state) are interesting routes of tuning the band gap width and character of the electronic structure. 
Properties of Eu-based materials from this family can be further affected by application of the external pressure, which can result in: (i) increase of magnetic ordering temperatures, (ii) change of the type of magnetic order (antiferromagnetic to ferromagnetic), (iii) change of the type of the crystal structure, or (iv) insulator to metal transition~\cite{Yu2020, Luo2023, Chen2023}.

Recently  one of such compounds, \EZP{}, in which spin magnetic moments of Eu$^{2+}$ ions ($S = 7/2$, $\mu_S = 7\ \muB$) interact to form an antiferromagnetic state below the N\'{e}el temperature $\TN$ = 23 K, has been intensively investigated  \cite{Berry2022, Singh2023, Krebber2023}. 
Its magnetic moments were assumed to order ferromagnetically in the basal-plane sheets with \Euion{} moments perpendicular to the \cax, and the coupling of the neighbor sheets being weaker antiferromagnetic~\cite{Berry2022}. 
The stronger ferromagnetic (F) interaction brings a positive paramagnetic Curie temperature $\theta_p = \SI{41.9}{K}$ recorded for field along the \cax. 
Low electrical conductivity and its thermally activated behavior demonstrates a semiconducting behavior with a band gap of 0.11 eV. 
Based on the absence of conduction electrons needed for Ruderman–Kittel–Kasuya–Yoshida (RKKY) exchange interaction, it was argued that the relatively high $\TN$ is due to a dipolar interaction. The fact that $\TN$ is the highest among the known $\ce{EuZn2$X$2}$ compounds (where $X$ = P, As, Sb, see \cite[and citations therein]{Berry2022}) is given in the context of the smallest lattice volume.  The suggested dipole-dipole interaction is, however, to our view unlikely to yield $\TN$ exceeding $T = \SI{20}{K}$.  
 Moreover, the conductivity was not studied to low enough temperatures  to show its possible interplay with the magnetic order. 

In Ref.~\cite{Singh2023} exchange interactions were analyzed and it was concluded 
that the superexchange mechanism including the P states is the plausible source of the ferromagnetic coupling within the basal-plane sheets.

In the more recent study \cite{Krebber2023} it was found that the Eu moments indeed order ferromagnetically in layers with the ($+ - + -$) coupling along the \cax, but they are tilted from the basal plane by approx. $\SI{40}{\degree}$. 
It was also shown that a colossal magnetoresistance effect (CMR) was observed from the ordered region to temperatures far exceeding the $\TN$. 

Our work concentrates on thorough investigation of structural, magnetic, and electronic properties of single crystals at ambient conditions as well as under hydrostatic pressure, which gradually increases the N\'{e}el temperature and reduces the electron gap width.
The $^{151}$Eu M{\"o}ssbauer spectroscopy and magnetic properties measurements at ambient pressure confirm that the Eu moments are indeed inclined from the basal plane.
We also present results of \textit{ab initio} calculations of the lattice dynamics and their implications for the heat capacity analysis. 

\section{Methods}
Single crystals of EuZn$_2$P$_2$ were grown by a Sn flux method in analogy with the process used for EuZn$_2$As$_2$~\cite{Bukowski2022}. Typical size of the crystals were $2\times2\times 0.2 \ \si{mm}$ (see Supplemental Materials \cite{SuppMatt}). Powder X-ray diffraction (XRD) measurements were made by a Panalytical Empyrean diffractometer. 
For the low-temperature XRD studies an Oxford Instruments PheniX closed-cycle helium refrigerator was used (14-300 K). 
During the measurements the sample position was corrected against thermal displacement of the sample stage. 
The stabilization of temperature was better than 0.1 K. 
The XRD patterns were refined using the Rietveld-type package FullProf~\cite{Rodriguez1993}.

The dc magnetization and ac magnetic susceptibility were measured in the temperature range 2–\SI{300}{K} in external magnetic fields $\muH$ up to \SI{9}{T} using VSM or ACMS option of a Quantum Design Physical Property Measurement System (PPMS-9). The ac susceptibility was measured using the ac driving field of \SI{1}{mT} with frequency of \SI{1111}{Hz}. In none of the methods we applied any demagnetization field correction, which could be difficult due to the irregular sample shapes. However, one should consider that the demagnetization field can have a substantial impact for the $H\parallel c$ geometry.  

The heat capacity measurements were carried out by a two-tau relaxation method with the heat capacity options of a Quantum Design Physical Property Measurement System (PPMS-9) in the temperature range 1.85-296 K. The sample for heat capacity measurements was attached and thermally coupled to the addenda with Apiezon N grease. A background signal from addenda and grease (versus \textit{T}) was recorded, as well.

The temperature dependence of electrical resistivity $\rho(T)$ at ambient pressure was measured in the PPMS cryostat (Quantum Design) with Keithley 6517 electrometer in the  Alternating Polarity Resistance mode. The temperature range 2 - 300 K was covered in zero magnetic field. In addition, measurements in applied magnetic fields of \SI{0.1}{T} and \SI{1}{T} were performed at low temperatures. Electrical contacts were realized by $\SI{25}{\mu m}$ AlSi wires attached by ultrasonic bonding (HB10 TPT wire bonder). Contact points were arranged within the $ab$-plane in a straight line with a middle point connected to source high output and the outer contacts kept at low potential (ground). The sample resistance in the measured temperature and field range changed by several orders of magnitude, which required changes of the electrometer voltage source settings. The high temperature region (above 135 K) was measured also by means of the Quantum Design Electrical Transport Option. 

\begin{figure}[!ht]
\centering
\includegraphics[width=0.90\linewidth]{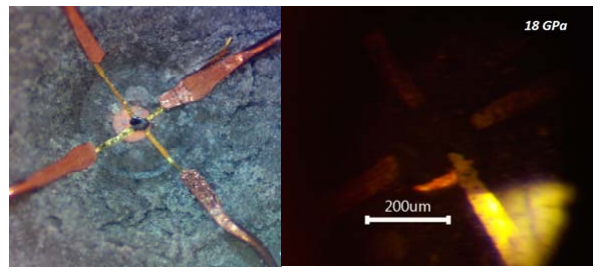}
\caption{Arrangement of the contacts for high-pressure resistivity measurements in DAC (left). The picture on the right shows the enlarged detail of the sample at $p = \SI{18}{GPa}$.}
\label{Pressure_cell1}
\end{figure}

The $\rho(T)$ measurements under applied pressure up to \SI{3}{GPa} were accomplished by means of the Electrical Transport Option with colloidal silver paste and gold wires providing electrical contacts. The geometrical arrangement was equivalent to the ambient pressure measurement. The double-wall pressure cell with a nominal maximum pressure of \SI{3}{GPa} \cite{Fujiwara2007} was used to apply the hydrostatic pressure. The Daphne oil 7474 \cite{Murata2008, Stasko2020} was used as the pressure transmitting medium. The pressure was determined at room temperature from the change of resistance of a thermally stabilized manganin wire. To minimize the thermal gradient during the measurement with pressure cell the rate of temperature change was \SI{0.5}{K/min} and all curves were measured as zero field cooled (ZFC). The pressure range above \SI{3}{GPa} was probed using diamond anvil cell (DAC) with NaCl quasi-hydrostatic pressure transmitting medium. The pressure was measured by shift of ruby fluorescence and the electrical contacts were realized by four gold foils arranged in Van der Pauw configuration (see~Fig.~\ref{Pressure_cell1}). This arrangement allowed us to calculate the absolute resistivity in the final high pressure metallic state. The great attention was paid to careful orientation of the samples. After orientation by X-ray diffraction the sample for DAC measurement was polished to 10 µm and piece of approximate size of $\SI{200}{\mu m} \times \SI{150}{\mu m}$ was inserted into the sample space of CuBe gasket insulated on one side by alumina filled epoxy.

Magnetization under high pressures has been measured in a SQUID magnetometer (MPMS XL, Quantum Design) using miniature diamond anvil pressure cell made of non-magnetic CuBe alloy \cite{giriat_turnbuckle_2010}. Daphne 7474 has been used as the pressure transmitting medium, with a ruby as a manometer. A microgram quantity of the sample has been used with approximate orientation of field parallel to the \cax. 
Measurement of empty cell without sample has been used as a reference to subtract the background signal of the pressure cell.

$^{151}$Eu M\"ossbauer spectra were recorded at \SI{300}{K}, \SI{30}{K}, and \SI{4.2}{K} with a conventional constant acceleration spectrometer using the $\ce{{}^{151}Sm(SmF3)}$ source. 
The \SI{21.5}{keV} $\gamma$-rays were detected with a NaI(Tl) scintillation detector.
The absorber surface density was approx. $\SI{60}{mg/cm^2}$. 
The spectra were analyzed by means of a least squares
fitting procedure. 
The absorption line positions and relative intensities were calculated by numerical diagonalization of the full hyperfine interactions Hamiltonian.
The isomer shift $\delta$ is given relative to the $\ce{^{151}Sm(SmF3)}$ source at room temperature

\section{Results and discussion}

\subsection{XRD and SEM}
\begin{figure}[!ht]
\centering
\includegraphics[width=0.95\linewidth]{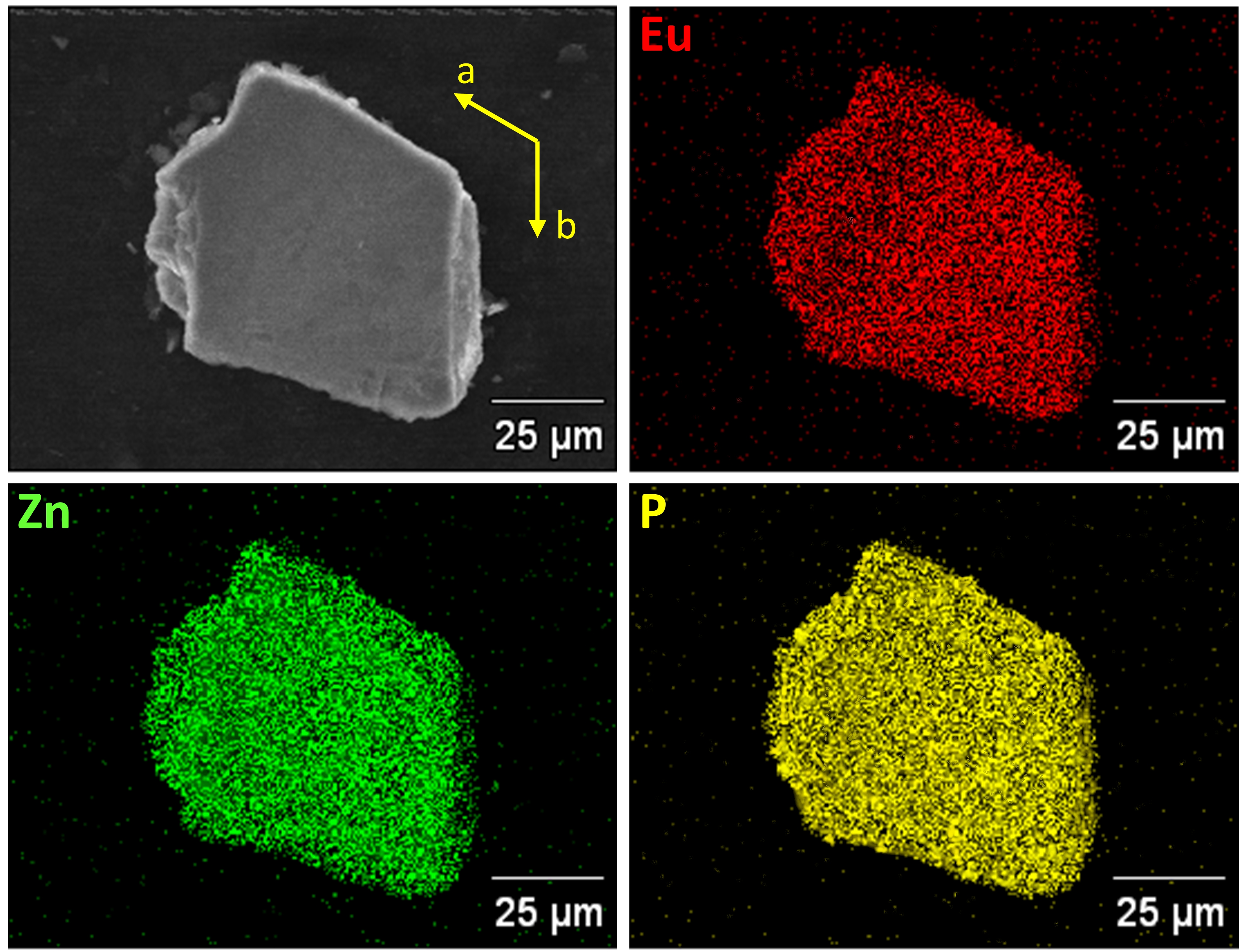}
\caption{SEM micrograph of a selected EuZn$_2$P$_2$ crystal with individual elemental maps provided by x-ray microanalysis.}
\label{SEM}
\end{figure}

Microstructure and composition of the EuZn$_2$P$_2$ single crystals were studied by scanning electron microscopy (SEM). An example is shown in Fig. \ref{SEM}. The elemental distribution is homogeneous, as checked for few large crystals and several small pieces. 
The overall elemental abundance Eu 19.4 at.\%, Zn 39.6 at.\%, P 41.0 at.\%, was found, which is in a good agreement with theoretical stoichiometry. 
The trigonal structure was clearly imprinted onto the crystals morphology with the large flat surfaces representing the basal planes, as checked by X\nobreakdash-ray diffraction.

\begin{figure}[!ht]
\centering
\includegraphics[width=1\linewidth]{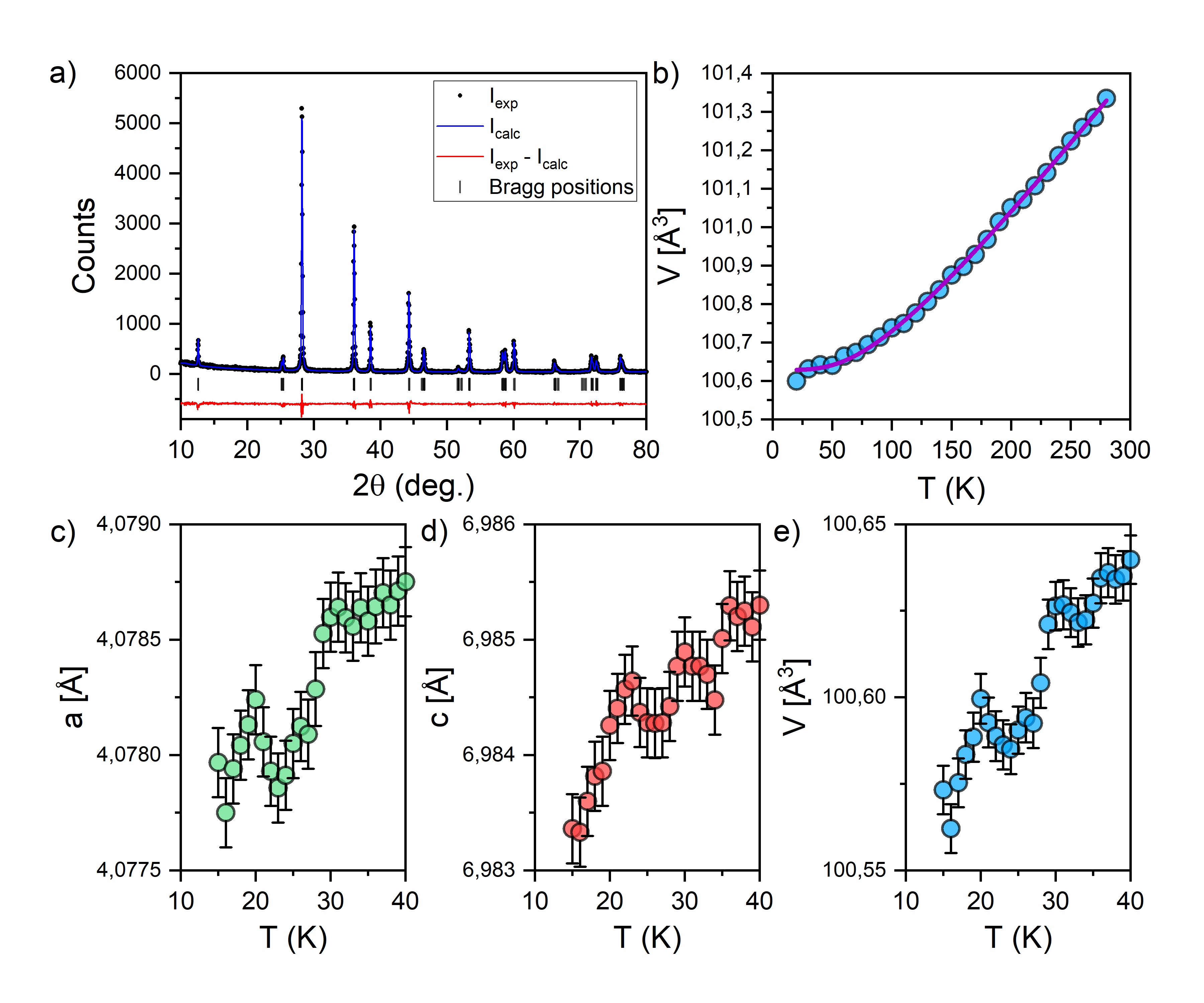}
\caption{Results of X-ray diffraction studies in the temperature range 15-300 K: Rietveld refinement of diffraction pattern collected at 15 K (a); unit cell volume in entire experimental range with fitted Debye curve (b); low temperature behavior of individual lattice parameters (c,d) and unit cell volume (e).}
\label{XRD}
\end{figure}

X-ray diffraction (XRD) patterns of EuZn$_2$P$_2$ collected in the temperature 15-300 K range can be fully indexed on the basis of the trigonal  $P\bar{3}m1$ space group (No. 164). 
The refined XRD pattern at $T = \SI{15}{K}$ is shown in Fig.~\ref{XRD}.
The refinement and absence of any spurious phase indicate a good quality of the sample.  
The lattice parameters obtained are $a = \SI{4.0779(1)}{\angstrom}$, $c = \SI{6.9834(3)}{\angstrom}$ and $a = \SI{4.0866(1)}{\angstrom}$, $c = \SI{7.0066(3)}{\angstrom}$ at $T = 15$ and \SI{300}{K}, respectively. 
The values at \SI{300}{K} are in a good agreement with the previously reported structure data~\cite{Berry2022}. 
The most interesting structure features were broadly discussed in our earlier paper concerning the isostructural system \ce{EuZn2As2}~\cite{Bukowski2022}.

The temperature dependence of the unit cell volume, presented in Fig.~\ref{XRD}(b), could be in the paramagnetic state approximated by the Debye formula\cite{Sayetat1998}:
\begin{equation}
 V = V_0 + I_C \frac{T^4}{\theta ^3 _D} \int\limits_{0}^{\frac{\ThD}{T}} \dfrac{x^3}{e^x - 1} \,dx,    
\end{equation}
where $V_0$ is the unit cell volume extrapolated to 0 K, $I_C$ is a slope of the linear part of the $V(T)$ dependent on Gr{\"u}neisen and compressibility parameters, while $\ThD$ is the Debye temperature. Both $a(T)$ and $c(T)$ (see Supplemental Materials \cite{SuppMatt}) are linear above approx. 130 K. This allows to evaluate individual linear thermal expansion coefficients $\alpha_a = 1.01\cdot \si{10^{-5}\ K^{-1}}$ and $\alpha_c = 1.48\cdot \si{10^{-5}\ K^{-1}}$, which gives the volume thermal expansion $\alpha_V = 3.50\cdot \si{10^{-5}\ K^{-1}}$.
The extrapolated unit cell volume was found to be $V_0 = \SI{100.63(1)}{\angstrom^3}$, the $I_C = \SI{0.0119(1)}{\angstrom^3 K^{-1}}$, and the Debye temperature $\ThD = \SI{324(11)}{K}$. 
The Debye temperature is higher than for the analogous arsenide EuZn$_2$As$_2$ (261 K)~\cite{Bukowski2022}, which can be attributed to the difference in atomic mass,  leading to the increase of the  phonon velocity. At low temperatures, below 40 K, an anomalous behavior of lattice parameters can be related to magnetic phenomena, see Figs. \ref{XRD}(c-e). 
Namely, both lattice parameters exhibit a sudden drop at $T = \SI{30}{K}$ for the $a$ parameter, and at 24 K for $c$. 
The $a$ parameter has a maximum at \SI{20}{K}, followed by the next drop below \SI{19}{K}. 
These features are also reflected in $V(T)$ (Fig.~\ref{XRD}(e) as two steps at 30 K and 20 K, respectively. 
Throughout all the $T$ range, the internal structure parameters determining the position of the Zn and P atoms are invariable within the experimental uncertainty. 

\subsection{Heat capacity measurements}

\begin{figure}[!ht]
\centering
\includegraphics[width=0.95\linewidth]{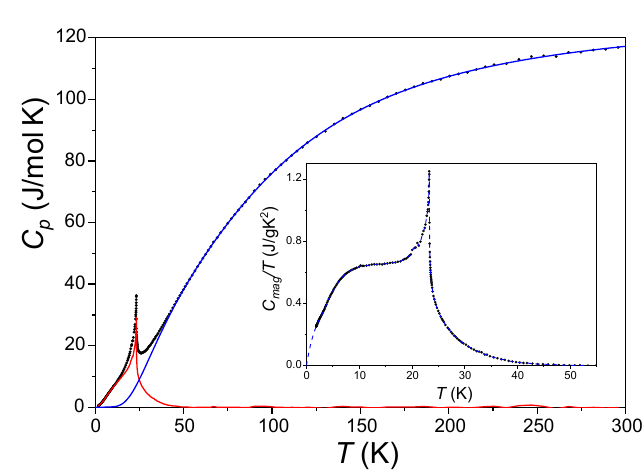}
\caption{Temperature dependence of the measured specific heat $C_p$ (black). The solid blue line is a fit by Eq. (\ref{eq:Cp}) and the solid red line is the calculated  magnetic contribution $\Cmag$ to the heat capacity. The inset shows $\Cmag/T$ as a function of \textit{T}. The dashed line in the inset is an extrapolation of the data to $T = \SI{0}{K}$ and the guide to the eye at higher $T$.}
\label{heat1}
\end{figure}

Temperature dependence of the specific heat ($C_p$) as well as its  magnetic part ($\Cmag$) are shown in Figure \ref{heat1}. 
One can notice a sharp peak at $T = \SI{23.3}{K}$, which reflects critical temperature of the antiferromagnetic ordering (N\'{e}el temperature $\TN$) of the \Euion{} moments. Due to the absence of crystal-field phenomena for spherically symmetric $4f^7$ state, no magnetic entropy should be present sufficiently far above  $\TN$. In the absence of electronic heat capacity, the paramagnetic $C_p(T)$ is given by the lattice contribution. In the simple approach we approximate it using a convenient mix of acoustic and optical phonons, with characteristic temperatures obtained by guess and fit method. We assume that such function is valid even in the magnetic state, which allows us to extract the magnetic heat capacity and magnetic entropy. In particular, we used the expression: 
\begin{widetext}
\begin{equation} \label{eq:Cp}
\Cph=\dfrac{R}{1-\alpha T} \left[9 \left( \dfrac{T}{\ThD} \right)^3 \int\limits_{0}^{\frac{\ThD}{T}} \dfrac{x^4 e^x}{ \left( e^x - 1 \right)^2} \,dx + \sum_{i} \frac{m_i \left( \dfrac{\ThEi}{T} \right)^2 e^{\ThEi/T} } {\left( e^{\ThEi/T}-1 \right)^2}\right],  
\end{equation}
\end{widetext}
where $\ThD$ is the Debye temperature, $\ThEi$ are Einstein temperatures with $m_i$ as corresponding multiplicities for each individual optical branch, $\alpha$ stands for an anharmonic coefficient, and $R$ is the universal gas constant~\cite{Ashcroft, Gondek2007}. 
In the recent reports~\cite{Berry2022, Krebber2023} a model with two Debye temperatures was proposed, but our model describes the experimental data better.
In order to facilitate the analysis, the summation over 12 independent optical branches was grouped into 2 branches with 6-fold degeneracy. 
The fit to the experimental data was done in the temperature range from \SI{48}{K} to \SI{296}{K}, i.e.,  far from the $\TN$, and it is represented by the solid blue curve in Fig.~\ref{heat1}.
The obtained fiting parameters are summarized in Table~\ref{Tab1}.
One can notice that the Debye temperature obtained, $\ThD = 277\pm\SI{8}{K}$, is in a reasonable agreement with estimate obtained from the lattice expansion ($324\pm\SI{11}{K}$).  
The magnetic part of the specific heat 
\begin{equation}
\Cmag(T) = C_p(T) - \Cph(T)   
\end{equation}
is shown in Fig.~\ref{heat1} as a solid red line.
As can be seen, $\Cmag$ remains finite in a limited range above $\TN$ and vanishes only at temperatures above $\approx\SI{50}{K}$, which is is a sign of short range magnetic correlations \cite{Pakhira2020, Pakhira2023}.

\begin{table}[!ht]
\caption{Debye ($\ThD$) and Einstein ($\ThEi$) temperatures and anharmonic coefficient ($\alpha$) obtained from the fitting of Eq. (\ref{eq:Cp}) to the temperature dependence of $C_p$.}
\label{Tab1}
\begin{tabular}{c| c} 
 \hline
 $\ThD$ (K) & $277 \pm 8$  \\ \hline
 $\theta_{\text{E}_1}$ (K) & $130\pm 2$  \\ \hline
 $\theta_{\text{E}_2}$ (K)& $406\pm 2$  \\ \hline
 $\alpha$ (1/K)& $(3.6\pm 0.3)\cdot 10^{-5}$  \\ \hline
\end{tabular}
\end{table}

The magnetic entropy $\Smag(T)$ was calculated from the $\Cmag(T)$ using the following formula:
\begin{equation} \label{eq:Sm}
\Smag(T)=\int\limits_{0}^{T} \dfrac{\Cmag}{T'} \,dT'. 
\end{equation}
As the measurement was carried out only down to \SI{1.85}{K}, the $\Cmag/T$ data were extrapolated from \SI{1.85}{K} to \SI{0 }{K} using a third-order polynomial function (fitted to the low-temperature experimental points). 
$\Smag$ calculated between $T = \SI{0}{K}$ and \SI{50}{K} was determined by numerical integration of the dashed curve shown in the inset in Fig.~\ref{heat1} and is equal to \SI{16.4}{J.mol^{-1} K^{-1}}.  
The expected value of $\Smag$ of one mole of particles with spin $J$ in a magnetic field is given by $\Smag = R \ln{(2J+1)}$, which for $J=7/2$ (value for Eu$^{2+}$ ions) amounts to \SI{17.28}{J.mol^{-1} K^{-1}}. 
Our experimental value is somewhat lower than the theoretical one, but much closer than previously reported one (\SI{11.3}{J.mol^{-1} K^{-1}})~\cite{Berry2022}, which is most likely owing to better estimate of the phonon contribution in our model compared to the two-Debye model proposed in Ref.~\cite{Berry2022}.

\begin{figure}[!ht]
\centering
\includegraphics[width=0.95\linewidth]{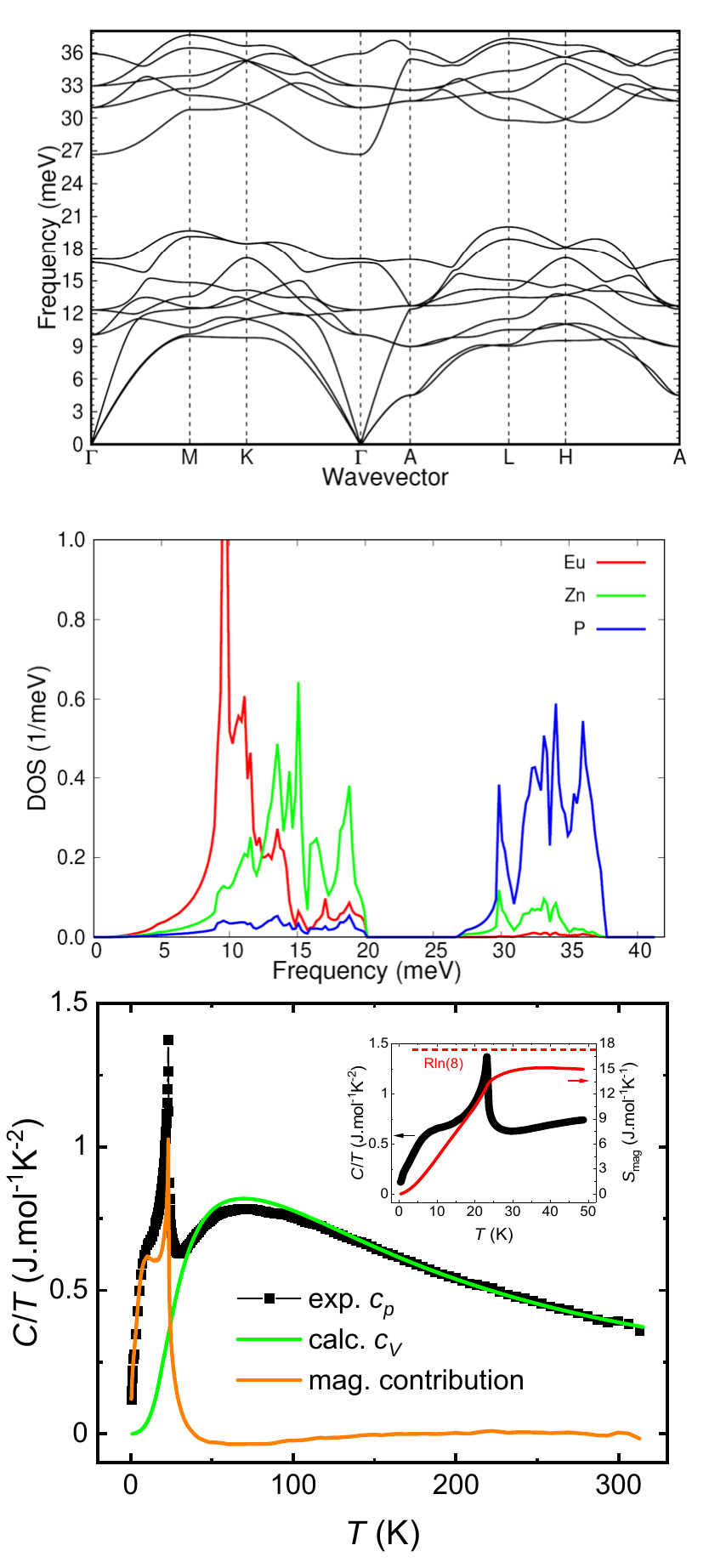}
\caption{Calculated phonon dispersion relations (top) and partial phonon densities of states decomposed into contributions of individual elements (middle panel). The lower panel displays the experimental specific heat $C_p$ data together with theoretical phonon specific heat calculated for constant volume, $C_V$ (green line), used to separate the magnetic contribution $\Cmag$ (orange line) and magnetic entropy (red line), shown in the inset.}
\label{heat_phonons2}
\end{figure}

It is interesting to compare such empirical approach (with multiple tunable parameters) with a fundamental parameter-free approach based on \textit{ab initio} calculations of lattice dynamics, which describe phonon dispersion relations including a dispersion in optical modes. Fig. \ref{heat_phonons2} shows the calculated phonon dispersion relations and respective densities of phonon states. One can distinguish dominant Eu-based acoustic modes at the lowest energies, followed by Eu optical modes at 10-\SI{13 }{meV}. The Zn optical modes extend up to \SI{20 }{meV}. High energy optical modes related to P vibrations are concentrated into the range  30-\SI{37 }{meV}. A careful examination reveals an inflexion point in the U-based phonon DOS at \SI{4 }{meV}. Analysis of individual dispersion relations reveals that below this energy there is rather soft transversal acoustic Eu phonon mode from the $\Gamma$ to A point of the Brillouin zone, i.e., the mode with the wavevector $q$ along the \cax{}. As a whole, the phonon spectrum is strongly non-Debye like. 

Calculated phonon density of states (DOS) allows to determine a theoretical lattice heat capacity at constant volume, $C_V$, displayed in the bottom panel of Fig.~\ref{heat_phonons2} as the green line. The remaining non-phononic part of the heat capacity we attribute to the magnetic  heat capacity, $\Cmag$, and it can be again taken to estimate the magnetic entropy, which is calculated as \SI{15.2}{J mol^{-1} K^{-1}}. Such value is somewhat lower than the previous estimate and hence more below the theoretical limit. Fig.~\ref{heat_phonons2} reveals a small issue of the lattice specific heat in the 50-\SI{80 }{K} range. The experimental data are slightly below the theoretical curve. Above this range the theory and experiment coincide. It suggests that the computations, bound to the $T = 0$ limit, may overestimate the softness of the lattice at low temperatures, which would reduce the calculated magnetic entropy in the ordered state. Nevertheless the theoretical phonon DOS gives a justification to the chosen Einstein temperatures given in Table~\ref{Tab1}. The $\theta_{\text{E}_1}$ =  $\SI{130}{K}$ corresponds to \SI{11.2 }{meV}, roughly in the middle of the energy range of Eu optical phonons, while  $\theta_{\text{E}_1}$ =  $\SI{406}{K}$, i.e., \SI{35 }{meV}, stands for energies of P optical phonons.

\begin{figure}[!ht]
\centering
\includegraphics[width=1\linewidth]{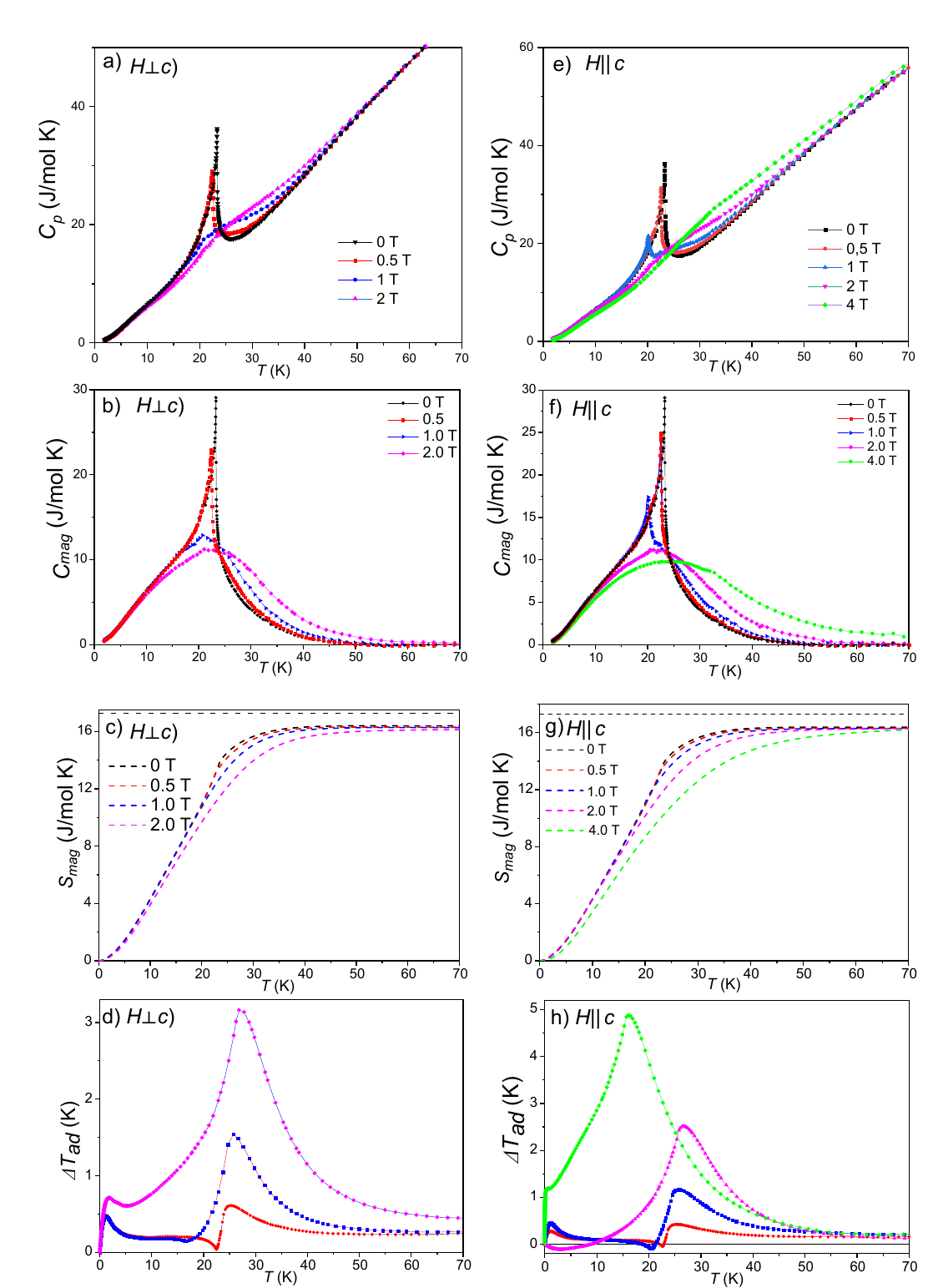}
\caption{Temperature and magnetic field dependence of the measured specific heat $C_p$, for magnetic fields perpendicular and parallel to the $c$ axis in panel (a) and (e), respectively. Panels (b) and (f) show the magnetic contribution $\Cmag$ to heat capacity. Panels (c) and (g) show  magnetic entropy $\Smag$ calculated from the data in the (b) and (f) panels, black dashed lines for $\Smag=\SI{17.28}{J/mol\cdot K}$ indicate a theoretical value of $\Smag$. Panels  (d) and (h) show adiabatic temperature changes $\Delta \Tad$ as a function of temperature.}
\label{heat2}
\end{figure}

Fig.~\ref{heat2} shows results of measurements of the heat capacity in magnetic fields, applied parallel to the \cax{} or perpendicular to it.
The sharp peak indicating the second-order phase transition to the antiferromagnetic state, which is clearly visible in zero field, shifts towards lower temperatures, becomes broader and eventually vanishes at fields exceeding the critical field needed for ferromagnetic-like alignment of Eu moments. Higher magnetic fields shift the magnetic entropy towards higher temperatures, in analogy with common ferromagnets.    
The critical field where the phase transition disappears takes place at lower field for the $H \perp c$ orientation, which is in agreement with the $M(H)$ curves, where the saturation of magnetisation (see Fig.~\ref{fig:Mag}) happens in lower fields. 
The behavior of the magnetic contribution $\Cmag$, which was calculated after subtracting the lattice contribution approximated with fitted theoretical curve, Eq.~(\ref{eq:Cp}), is better visible in  Fig. \ref{heat2}(b) and (f) for both orientations.
The $\Cmag(T)$  shows features typical for antiferromagnetically ordered Eu compounds~\cite{Pakhira2020, Giovannini2021, Bukowski2022}, i.e. it shifts towards low temperatures and broadens.
The exact temperatures of the peak related to the antiferromagnetic order are shown below in Fig.~\ref{fig:PhDiag}.

We also calculated the magnetic entropy $\Smag$ using Eq.~(\ref{eq:Sm}) and its temperature and field dependencies are shown in panels (c) and (g) of Fig. \ref{heat2}.
For all fields and both orientations, $\Smag$ is lower than the theoretical value (\SI{17.28}{J mol^{-1} K^{-1}}) and it saturates at temperatures increasing with applied magnetic fields, reaching $T = \SI{70}{K}$ at our experimental conditions. 
The release of the entropy above $\TN$ due to short range  correlations was observed previously for other Eu compounds~\cite{Pakhira2020, Pakhira2021, Pakhira2022, Pakhira2023}. 
The collected data allow to calculate an adiabatic temperature change $\Delta \Tad$, using the relation:
\begin{equation} 
\Delta \Tad(T, \Delta H)=T(S, \Hfinal)-T(S, \Hinitial),
\label{eq:Ad}
\end{equation}
 where $T(S, H)$ is temperature as a function of the total entropy $S$ and magnetic field $H$ (initial $\Hinitial = 0$ and final $\Hfinal > 0$)~\cite{Gschneidner2005}.
Calculated values of $\Delta \Tad$ for both field orientations are presented in panels (d) and (h) of Fig.~\ref{heat2}.
For $\mu_0H = \SI{2}{T}$ perpendicular to the $c$ axis the maximum value of $\Delta \Tad$ amounts to \SI{3.16}{K} and is larger compared to \SI{2.52}{K} value obtained for the $H \parallel c$ orientation, indicating stronger magnetocaloric effect for the $H \perp c$ orientation.
For $H \parallel c$ at \SI{4 }{T} the maximum value of $\Delta \Tad$ almost doubles and amounts to \SI{4.88}{K}. 
All these observations are in line with the results obtained for very similar antiferromagnetically ordered EuZn$_2$As$_2$~\cite{Bukowski2022}. 
The maximum values of $\Delta \Tad$ are also comparable to those obtained for ferromagnetic EuO~\cite{Ahn2005}.

\subsection{Magnetic measurements}
\subsubsection{Ac susceptibility}
\begin{figure}
    \centering
    \includegraphics[width=0.9\linewidth]{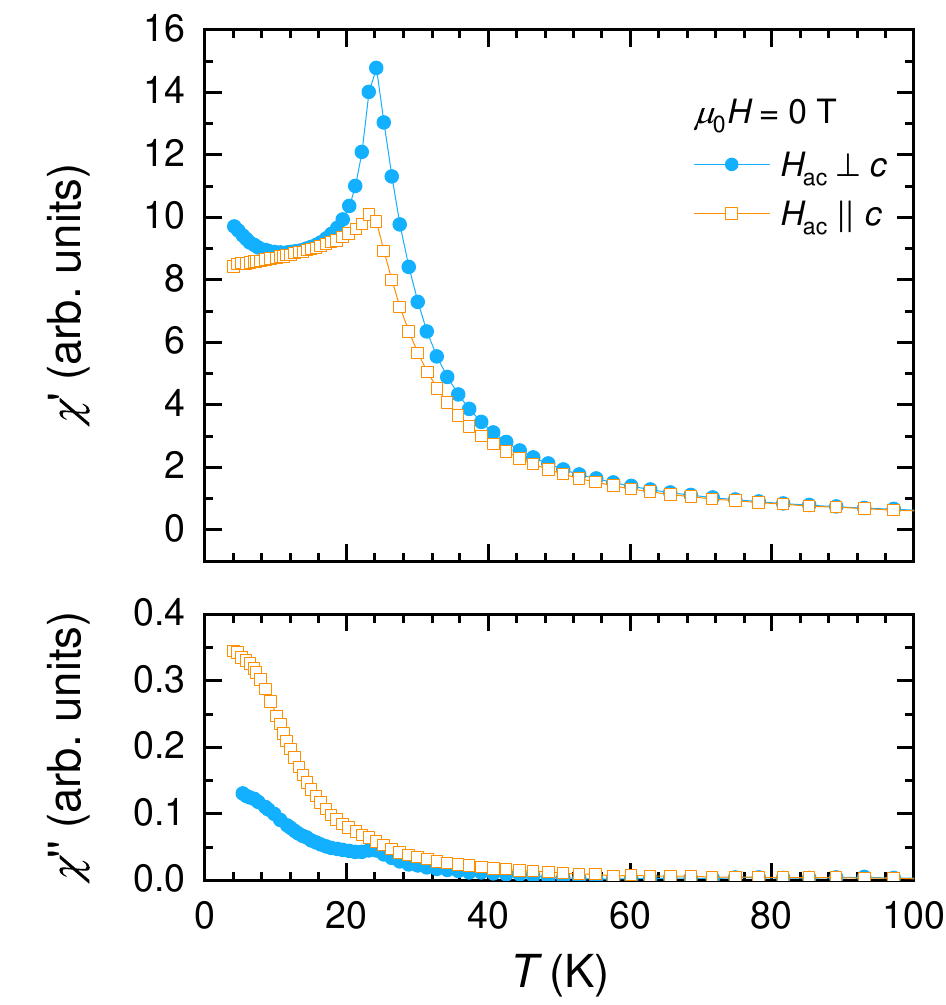}
    \caption{Real (top) and imaginary (bottom) parts of ac susceptibility investigated with driving field of \SI{1}{mT} applied perpendicular (closed blue circles) and parallel (open orange squares) to the $c$ axis measured in zero static magnetic field.}
    \label{fig:AC0}
\end{figure}
\begin{figure}
    \centering
    \includegraphics[width=1\linewidth]{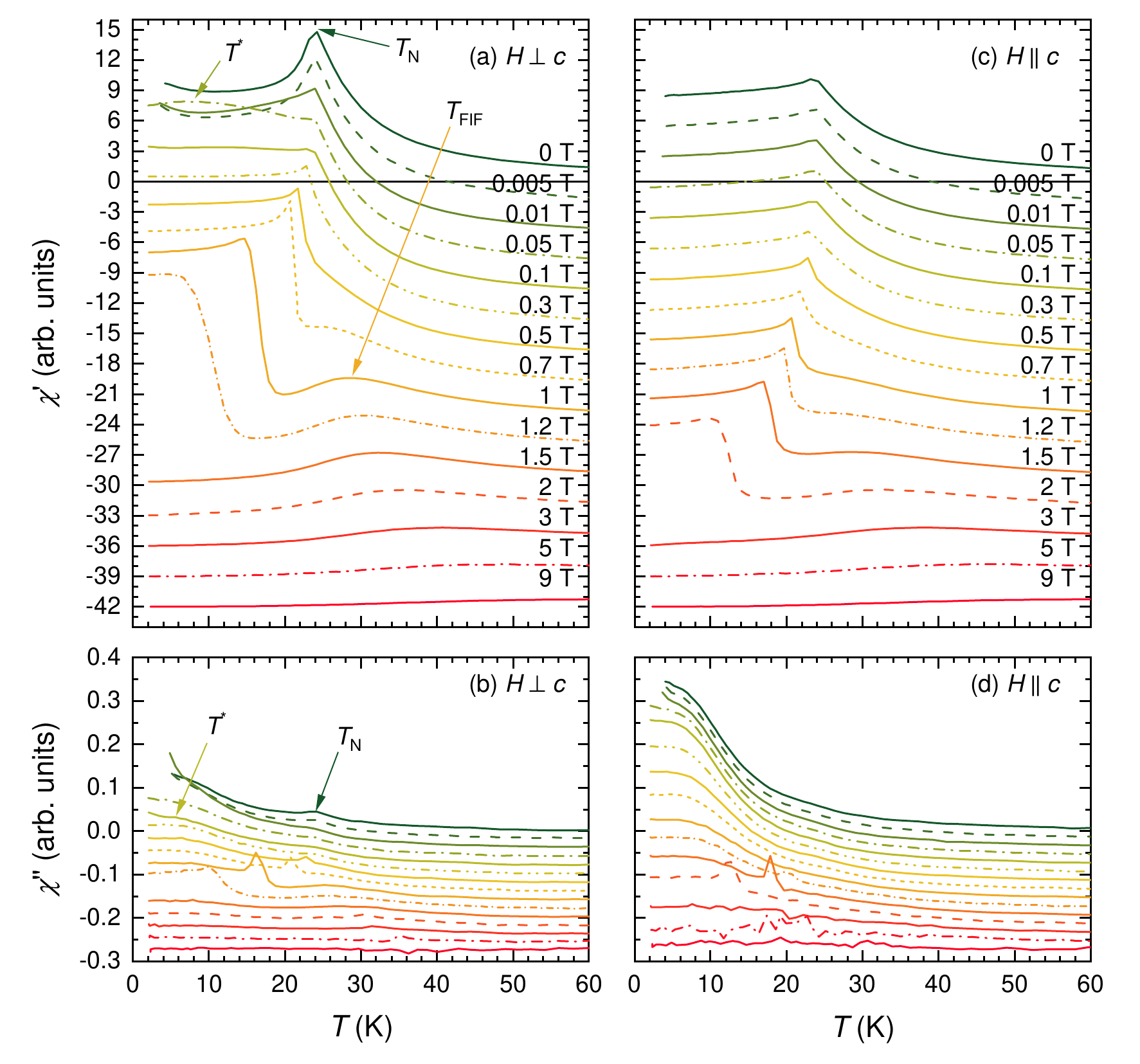}
    \caption{Real (a, c) and imaginary (b, d) parts of ac susceptibility investigated with driving field $\Hac=\SI{1}{mT}$  ($f = \SI{1111}{Hz}$) in various static magnetic fields $\muH$ applied perpendicular (a, b) and parallel (c, d) to the $c$ axis. The dependencies for $\muH\geq \SI{0.005}{T}$ are shifted along the y axis by 3 in panels (a, c) and by 0.02 in (b, d).}
    \label{fig:AC}
\end{figure}
Temperature dependence of ac susceptibility investigated in zero static magnetic field ($\muH=\SI{0}{T}$) using the ac driving field $\SI{1}{mT}$ applied perpendicular and parallel to the crystallographic \cax{} is shown in Fig.~\ref{fig:AC0}. As can be seen, the real part of susceptibility $\chi'(T)$ exhibits a sharp maximum at about \SI{23}{K}. At the same time a less pronounced maximum can be observed in the imaginary part of susceptibility $\chi''(T)$. 
As discussed below, these are the fingerprints of the antiferromagnetic transition. 

For a typical antiferromagnet with low magnetocrystalline anisotropy, one would expect below $\TN$ to see a decrease of susceptibility for the so called "easy-direction" and (nearly) constant values of susceptibility for the other direction. However, \EZP{} exhibits with decreasing temperature different behavior of $\chi'(T)$. The susceptibility increases for $\Hac\perp c$ and for $\Hac\parallel c$ it decreases first but a small kink is observed at about \SI{5}{K}.
In addition, the imaginary part of susceptibility $\chi''(T)$ increases (below $\TN$), which reflects energy dissipation  associated with remagnetization in the already magnetic state. 

For non-zero external static magnetic fields, additional anomalies are observed for the real and imaginary parts of ac susceptibility in both field directions (see Fig.~\ref{fig:AC}). While the maximum at $\SI{23}{K}$ (associated with the antiferromagnetic transition) shifts towards the lower temperatures with increasing $\mu_0H$ for both directions, the shape of $\chi'(T)$  depends on the direction. For $H \perp c$, $\chi'(T)$  below $\TN$ changes to nearly temperature independent (flat type), while for $H \parallel c$ it decreases with decreasing temperatures. For low fields ($\muH < \SI{0.3}{T}$ ) applied perpendicular to the \cax{},  $\chi'(T)$ exhibits a broad maximum, which is accompanied by a maximum in $\chi''(T)$ below the $\TN$, but for higher fields the maximum appears above $\TN$, in other words, this maximum is shifted towards higher temperatures with increasing static magnetic field. On the other hand, for $H \parallel c$ such maximum in $\chi'(T)$ is visible only for higher fields ($\SI{0.5}{T}$)
and the response in the $\chi''(T)$ is much more subtle than in the $H\perp c$  case. An anomaly that is shifted towards higher temperatures by increasing static magnetic field can be associated with a reorientation of \Euion{} magnetic moments towards the field direction. In static fields exceeding \SI{3 }{T}, a ferromagnetic-like alignment is achieved. Hence the ac susceptibility becomes low and practically featureless on the given temperature scale.

Similar behavior (increase for $\muH=\SI{0}{T}$  and nearly temperature independent $\chi(T)$ for $\muH>\SI{0}{T}$) of the real part of susceptibility below $\TN$ for $\Hac\perp c$  (and $H\perp c$) was observed for example for \ce{EuFe2As2} and \ce{Eu_{0.88}Ca_{0.12}Fe2As2} \cite{Tran2018,SuppMatt}. It was proposed that this is most likely due to the detwinning process, which we can not exclude in the case of $\EZP$. 

On the contrary, the $\chi'(T)$ dependence for $c\parallel \muH>\SI{0}{T}$ does not resemble susceptibility for the mentioned compounds (see Supplemental Materials \cite{SuppMatt}). We do find, however, some similarities for \ce{EuZn2As2} \cite{Bukowski2022}, as well as for the Co-doped \ce{EuFe2As2} compounds \cite{Tran2012NJP,Tran2023}.  For all of these compounds the magnetic moments of $\Euion$ are tilted from the basal plane. 

This can lead us to assumption that also in $\EZP$ the magnetic moments of $\Euion$  order antiferromagnetically but are tilted from the basal plane, in agreement with the M\"{o}ssbauer spectroscopy results presented below (see Section~\ref{sec:Moss}). Moreover, the observed ferromagnetic-like anomaly below the antiferromangetic transition can be associated with: (i) the detwinning process (or presence of antiferromagnetic domains), (ii) possible spiral order instead of an A-type antiferromagnetic one, which results in existence of an ferromagnetic component, i.e., ferromagnetic interactions between \Euion{} magnetic moments.  

\subsubsection{Dc magnetization}
Although the dc magnetization of \EZP{} was studied already \cite{Singh2023,Krebber2023,Berry2022}, 
we performed dc magnetization measurements in a broader range of temperatures and magnetic fields  for a more complete description. This also enabled us to draw a more detailed magnetic phase diagram of the investigated system (see Section~\ref{sec:PhDiag}).

\begin{figure}[!ht]
\centering
\includegraphics[width=0.49\linewidth]{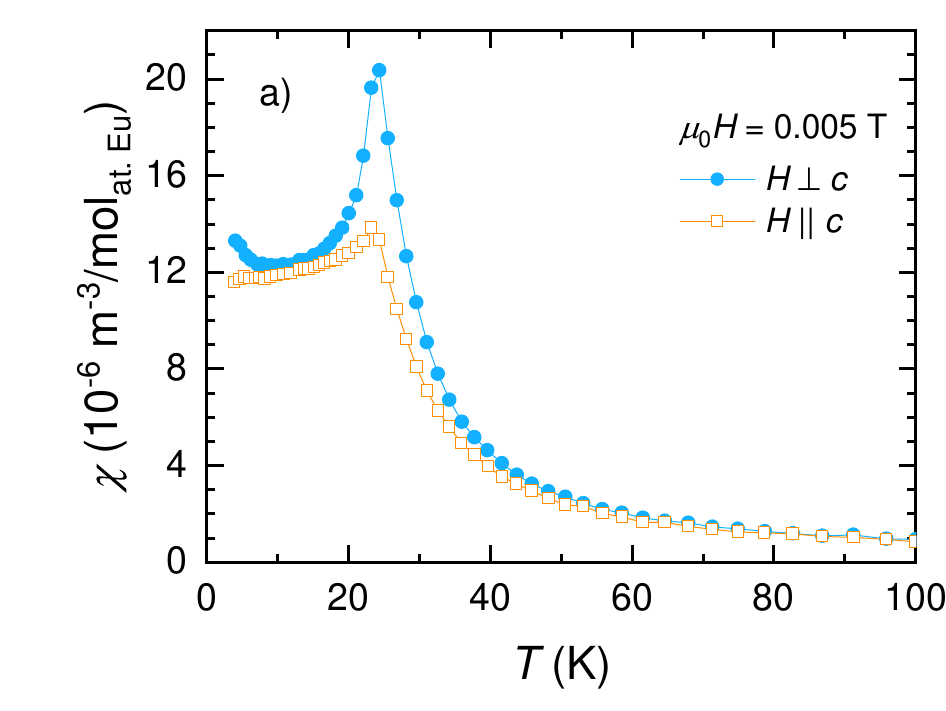}
\includegraphics[width=0.49\linewidth]{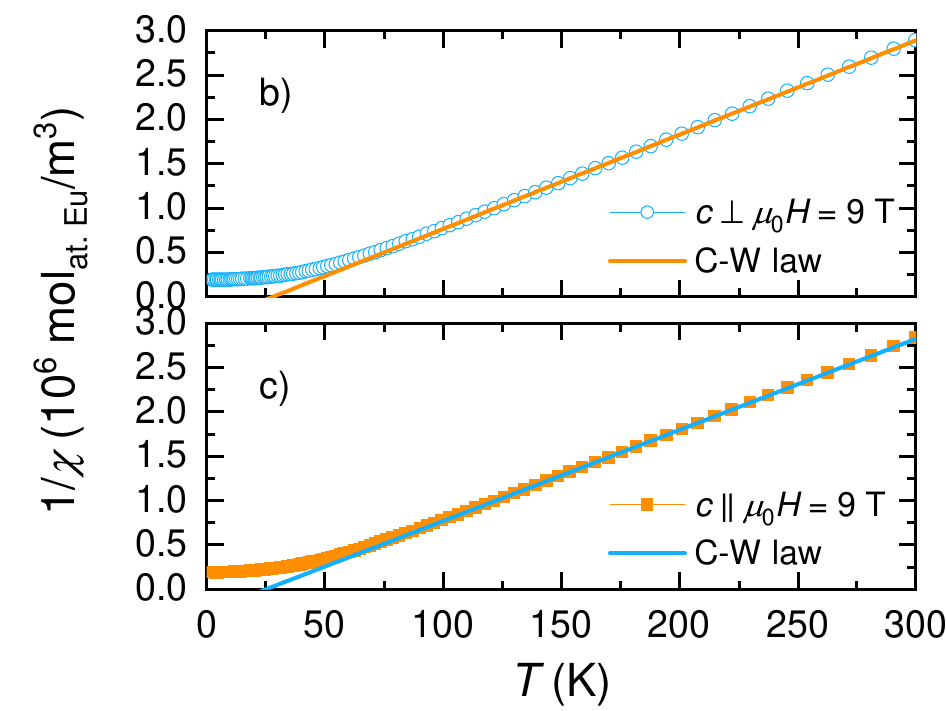}
\caption{Temperature dependence of (a) dc magnetic susceptibility $\chi$ investigated in $\muH = \SI{5}{mT}$ applied perpendicular (closed blue circles) and parallel (open orange squares) to the crystallographic $c$ axis, and (b, c) inverted susceptibility $1/\chi$ measured in $\muH = \SI{9}{T}$  applied perpendicular (b, open blue circles) and parallel (c, closed orange squares) to the crystallographic $c$ axis together with respective fits with the Curie-Weiss law, Eq.~\ref{eq:CW-law}, between $100$ and $\SI{300}{K}$ (solid line).}\label{Fig:DCSusc}
\end{figure}

Temperature dependence of dc magnetic susceptibility  investigated at  low magnetic fields applied  perpendicular ($H\perp c$) and parallel ($H\parallel c$) to the crystallographic \cax{}, strongly resembling the ac susceptibility data, is shown in Fig.~\ref{Fig:DCSusc}(a). It increases with decreasing temperatures until 
a kink at $\TN\approx\SI{23}{K}$ is reached, and then it decreases.

Temperature dependencies of the inverse susceptibility, $1/\chi$, measured in $\muH=\SI{9}{T}$ applied perpendicular and parallel to the \cax{} are shown in Fig.~\ref{Fig:DCSusc}(b) and (c), respectively. The $1/\chi(T)$ dependence is linear at high temperatures. Fitting the experimental data to the Curie-Weiss law:
\begin{equation}
\chi(T)=\dfrac{\NA}{3\kB\mu_0} \dfrac{\mueff^2}{T-\Thp},
\label{eq:CW-law}
\end{equation}
where $\NA$ is the the Avogadro number, $\kB$ the Boltzmann constant, $\mu_0$ permeability of vacuum, $\Thp$ the paramagnetic Curie temperature, $\mueff$ effective magnetic moment (in Bohr magnetons $\si{\micro_{B}}$) gives $\mueff^{\perp c} = (7.74 \pm 0.01)\ \si{\micro_B}$ and  $\mueff^{\parallel c} = (7.87 \pm 0.01)\ \si{\micro_B}$ for $H \perp c$ and $H \parallel c$, respectively; which is close to  $\SI{7.94}{\micro_B}$, the theoretical value for the $4f^7$  configuration. The evaluated paramagnetic Curie temperature, positive for both orientations ($\Thp^{\perp c} = (27.8 \pm 0.2)\ \si{K}$ and $\Thp^{\parallel c} = (24.8 \pm 0.3)\ \si{K}$) indicates that the ferromagnetic interactions are dominant despite the AF ground state. 
The above findings are in very good agreement with previous studies on this material \cite{Berry2022, Krebber2023, Singh2023}. The  anomaly at $\TN$ shifts towards lower temperatures (see Supplemental Materials~\cite{SuppMatt}) with increasing external magnetic field $H$, as usual in antiferromagnets.


\begin{figure}
    \centering
\includegraphics[width=0.49\linewidth]{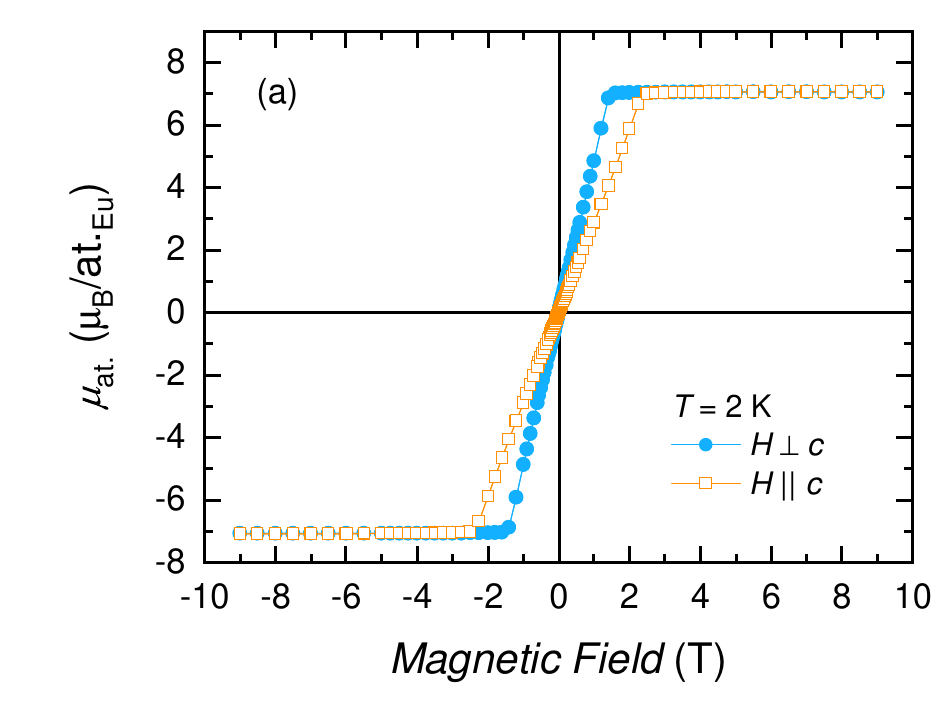}
\includegraphics[width=0.49\linewidth]{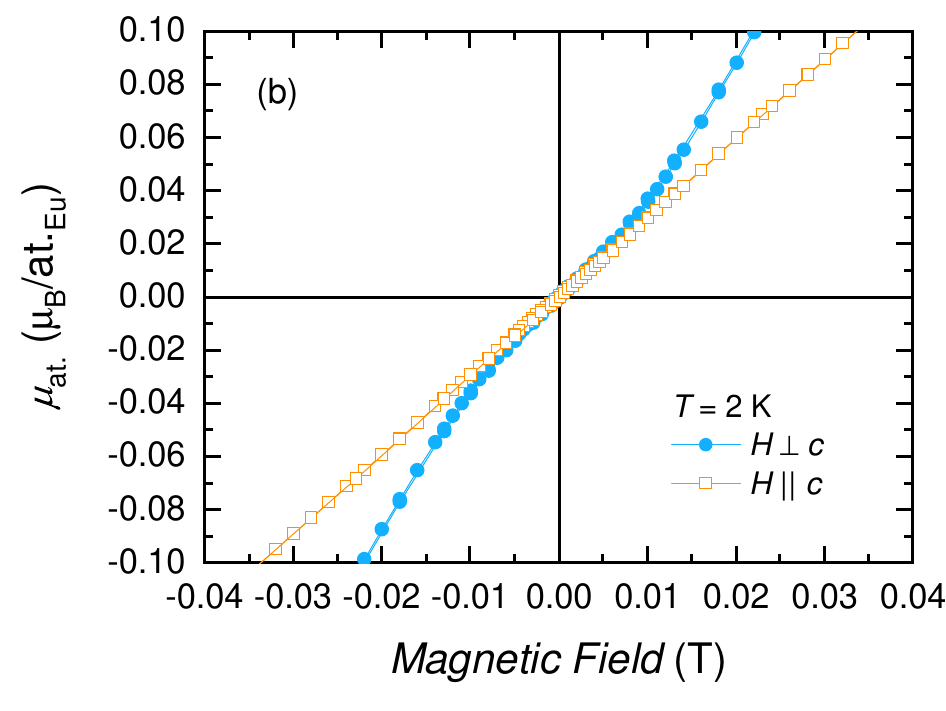}
\caption{
Magnetic field dependence of the dc magnetisation per \Euion{}  $\muat$ at \SI{2}{K} for $c\parallel H$ and $c\perp H$ orientations (a) for magnetic fields up to \SI{9}{T} with an enlarged view of the low field region (b).}
\label{fig:Mag}
\end{figure}
Field dependencies of magnetisation for $H \perp c$ and $H \parallel c$ orientations measured at \textit{T} = \SI{2}{K} are presented in Fig. \ref{fig:Mag}(a). 
Both curves show, as expected for an antiferromagnet, no spontaneous magnetization and no magnetic hysteresis. The response in low fields is linear. At sufficiently high fields  magnetisation for both orientations saturates to \SI{7}{\micro_{B}} per Eu atom, which is consistent with the theoretical value for \Euion{} ($\mu^\text{th}_\text{sat}= g J$, with $g=2$ and $J=7/2$).
The field at which saturation is reached depends on the orientation and is equal \SI{1.3}{T} and \SI{2.5}{T} for $H\perp c$ and $H\parallel c$ orientations, respectively. These fields correspond to a crossover to the field-induced ferromagnetic state, i.e., state at which all magnetic moments of \Euion{} align ferromagnetically under the influence of magnetic field.


At very low fields (below \SI{0.02}{T}),  see Fig.~\ref{fig:Mag}(b), one can notice that for the $H\perp c$ orientation there is a clear change of behavior of magnetisation.
It first increases slowly and then from about \SI{0.015}{T} it starts to increase more rapidly, maintaining the same slope until reaching the saturation.
It resembles the virgin magnetisation curve of the ferromagnets with domain wall pinning mechanism of the magnetisation reversal.
The same behavior was found in \EZP{} by \citeauthor{Krebber2023}~\cite{Krebber2023}, as well as in other Eu containing compounds  \cite{Zapf2014, Tran2018, Pakhira2022, Pakhira2023} and was attributed to either the presence of antiferromagnetic domains, which were vanishing at higher fields (above \SI{0.05}{T}), or the detwinning process.
This effect may be related to the presence of three equivalent directions for \Euion{} moments in the unit cell corresponding to the threefold rotational symmetry of the \cax{} \cite{Pakhira2023}.
Hence the ferromagnetic coupling is frustrated and the antiferromagnetic interaction can prevail.

\subsection{M{\"o}ssbauer spectroscopy measurements}\label{sec:Moss}

\begin{figure}[!ht]
\centering
\includegraphics[width=0.95\linewidth]{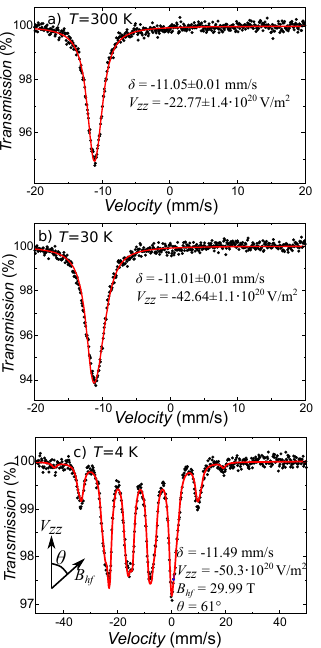}
\caption{$^{151}$Eu M\"ossbauer spectra of EuZn$_2$P$_2$ measured at (a) room temperature, (b) $T = \SI{30}{K}$ and  (c) \SI{4}{K}. Experimental data are shown as black squares and the red line is the result of the fit, which parameters ($\delta$, $V_{zz}$, $\Bhf$ and $\theta$) are also provided.}
\label{moss}
\end{figure}

There are very few experimental techniques, which can provide information regarding the orientation of ordered magnetic moments.
One of them is M{\"o}ssbauer spectroscopy and we carried out such measurements on the $^{151}$Eu isotope.
In Fig.~\ref{moss} we present results of measurements carried out in the ordered (at \textit{T} = \SI{4 }{K}) and paramagnetic state (\SI{30 }{K} and \SI{300 }{K}).
At the room temperature (Fig.~\ref{moss}(a)) the spectrum consists of a single absorption line with isomer shift $\delta=\SI{-11.09}{mm/s}$, which is typical for Eu$^{2+}$ ions \cite{Schellenberg2010, Blachowski2011, Komedera2021}.

The other parameters, which can be obtained from the fit of experimental data, are the largest component of the electric field gradient tensor (EFG), $V_{zz}$, and the quadrupole coupling, $\epsilon$, which is a result of the interaction of EFG with a nuclear electric quadrupole moment, $Q_g$.
EFG at the Eu site has two sources, the first one is due to its own electrons and the second comes from  neighbouring ions.
The Eu$^{2+}$ ion has a half-filled 4$f^7$ electronic configuration with the $^8$S$_{7/2}$ ground state, so that the 4$f$ electron contribution to the EFG is expected to be zero.
The second contribution coming from the "lattice" in our case is also expected to be rather small since the Eu site has a rather high symmetry (3$\overline{m}$ point symmetry).
Therefore, the EFG, $V_{zz}$ and the resulting quadrupole coupling, $\epsilon$, are expected to be small.
Since the \cax{} is a threefold rotation axis, one expects that it becomes the principal axis of the EFG tensor \cite{Yoshida} and the tensor itself is axially symmetric along it, i.e., the $V_{zz}$ component is along the \cax{}.
Then the quadrupole coupling $\epsilon$ in the paramagnetic state can be written as:
\begin{equation}
\epsilon= \frac{1}{4} \frac{c e Q_g V_{zz}}{E_{\gamma}},
\label{quad1}
\end{equation}
where $c$ is the light velocity in vacuum, $e$ electron charge, and $E_{\gamma}$ is the M\"ossbauer photon energy. 
In the case of $^{151}$Eu M\"ossbauer spectroscopy one can learn  the sign of $V_{zz}$ in the paramagnetic state from the asymmetry of the spectrum. 
The tail towards more negative (positive) values of the velocity indicates a positive (negative) sign of $V_{zz}$.
In our case $V_{zz}$ is small so it is difficult to distinguish its sign just by the eye. Therefore we calculated the skewness and determined its positive sign, meaning a negative sign of $V_{zz}$.  
From the fit to the data at $T = \SI{300}{K}$ we obtain $V_{zz} = -22.8\cdot 10^{20} \ \si{V/m2}$, which is comparable to the value for EuZn$_2$As$_2$~\cite{Bukowski2022} and is rather small~\cite{Albedah2018, Albedah2020}, as expected. 

In the magnetically ordered state the formula in Eq.~(\ref{quad1}) is modified as follows:
\begin{equation}
\epsilon= \frac{1}{4} \frac{c e Q_g V_{zz}}{E_{\gamma}} \left( \frac{3 \rm{cos}^2 \theta -1 }{2} \right),
\label{quad2}
\end{equation}
where $\theta$ is the angle between the principal component axis of the EFG ($c$ axis in our case) and the direction of Eu hyperfine field $\Bhf$ and, therefore, the Eu magnetic moment.
Hence it is in general possible to obtain information about the orientation of Eu magnetic moments. 
However, we note that $V_{zz}$ and $\theta$ are fitted simultaneously in the same term, Eq.~(\ref{quad2}), and more than one solution may provide similar fit quality.
In order to facilitate the fitting procedure in the ordered state, one could use parameters $\delta$ and $V_{zz}$ obtained in the paramagnetic state.
Since they are usually temperature dependent, we conducted a measurement just above $\TN$, i.e., at $T = \SI{30}{K}$, shown in Fig. \ref{moss}(b).
Compared to room temperature the isomer shift $\delta$ somewhat changes and the absolute value of $V_{zz}$ increases as is expected due to the lattice contraction and the corresponding decrease of inter-atomic distances~\cite{Torumba2006}.

We note here that there are three recent studies, which discuss possible orientation of Eu magnetic moments in EuZn$_2$P$_2$. \citeauthor{Berry2022} and \citeauthor{Singh2023} suggested that there is a planar A-type antiferromagnetic order, i.e., Eu moments lie in the basal plane \cite{Berry2022, Singh2023}, while \citeauthor{Krebber2023} using resonant x-ray diffraction reported that Eu moments are tilted from the basal plane by an angle of $\SI{40}{\degree} \pm \SI{10}{\degree}$~\cite{Krebber2023}.
At $T = \SI{4 }{K}$, the $^{151}$Eu spectrum consists of groups of peaks, which is due to the presence of a magnetic hyperfine field at the Eu nucleus, related to the magnetic order of Eu moments.
Below we discuss results of three fitting procedures with focus on $V_{zz}$ and $\theta$, since $\delta$ and $\Bhf$ are the same within the error margin (see Table~\ref{Tab2}).
They are also very similar to those obtained for EuZn$_2$As$_2$ \cite{Bukowski2022}.
First we performed a fit with all parameters free and we obtained $|V_{zz}|$ somewhat larger compared to that at $T = \SI{30 }{K}$, which is expected, and the angle $\theta = \SI{61}{\degree}$.
This fit is shown in Fig. \ref{moss}(c).
In the second attempt we fixed $V_{zz}$ at the value obtained at $T = \SI{30 }{K}$ and we obtained the angle $\theta=\SI{63}{\degree}$.
Obtained values of $\theta$ indicate that Eu moments are inclined from the basal plane by about $\SI{30}{\degree}$, in a good agreement with the value suggested by \citeauthor{Krebber2023}~\cite{Krebber2023}. 
In the final fit we fixed the angle $\theta$ at $\SI{90}{\degree}$, a value suggested by \citeauthor{Berry2022}, and we obtained $|V_{zz}|$ almost three times smaller compared to that at \SI{30}{K}, which is very unlikely, since $|V_{zz}|$ typically increases with decreasing $T$~\cite{Torumba2006, Albedah2018, Albedah2020}, unless there is a structure phase transition, or strong or uneven change of the lattice parameters with temperature. 
This is practically excluded by the XRD data discussed above, giving the $c/a$ ratio changes with $T$ only by 0.1\%, while $a$ and $c$ vary by 
0.2\% and 0.3\%, respectively. 
Therefore, our M\"ossbauer spectroscopy results indicate that the option with Eu moments in the basal plane is implausible and that they are tilted from the basal plane, as suggested by \citeauthor{Krebber2023} \cite{Krebber2023}. 
We also stress that our $M(H)$ curves in weak magnetic fields show the same behavior as in Ref.~\cite{Krebber2023}, while \citeauthor{Berry2022}~\cite{Berry2022} mentioned a linear behavior in weak fields.
In fact, such disagreement on the orientation of Eu moments is not unusual. 
It was observed by different groups studying the same compounds, for example, for structurally similar EuZn$_2$As$_2$~\cite{Bukowski2022, Blawat2022, Yi2023} or EuSn$_2$As$_2$~\cite{Li2019, Pakhira2021}.
The angle $\theta$ we found here for EuZn$_2$P$_2$ is similar to that in EuZn$_2$As$_2$~\cite{Bukowski2022} or EuNi$_2$As$_2$~\cite{Komedera2021}. 
Such inclination from the basal plane was indicated by \textit{ab initio} calculations assuming for simplicity a ferromagnetic coupling, giving equal Eu moment component on each Cartesian coordinate $x$, $y$, $z$. This yields the inclination from the basal plane $\SI{35}{\degree}$, i.e., $\theta=\SI{55}{\degree}$.  

\begin{table}[h!]
\caption{Parameters obtained from fitting of the spectrum obtained at $T= \SI{4}{K}$. In the first row all fitting parameters were free, in the second row $V_{zz}$ was fixed to its value at $T = \SI{30}{K}$, in the third row the angle $\theta$ was fixed at $\SI{90}{\degree}$, as suggested by \citeauthor{Berry2022} \cite{Berry2022}.}
\label{Tab2}
\begin{tabular}{|c| c| c| c|} 
 \hline
 $V_{zz} (\cdot10^{20} \rm{V/m^2})$ & $\theta (\rm{^o)}$ & $\delta$ (mm/s) &    $\Bhf$ (T) \\ \hline
-50.3 $\pm$ 5.1 & 61 $\pm$ 1 & -11.49 $\pm$ 0.01 & 29.99 $\pm$ 0.04 \\ \hline
-42.6 (fixed) & 63 $\pm$ 1 & -11.49 $\pm$ 0.01 & 29.97 $\pm$ 0.04 \\ \hline
-15.7 $\pm$ 1.6 & 90 (fixed) & -11.49 $\pm$ 0.01 & 29.83 $\pm$ 0.03  \\ \hline
\end{tabular}
\end{table}


\subsection{Magnetic phase diagram\label{sec:PhDiag}}
\begin{figure*}
    \centering
    \includegraphics[width=0.9\linewidth]{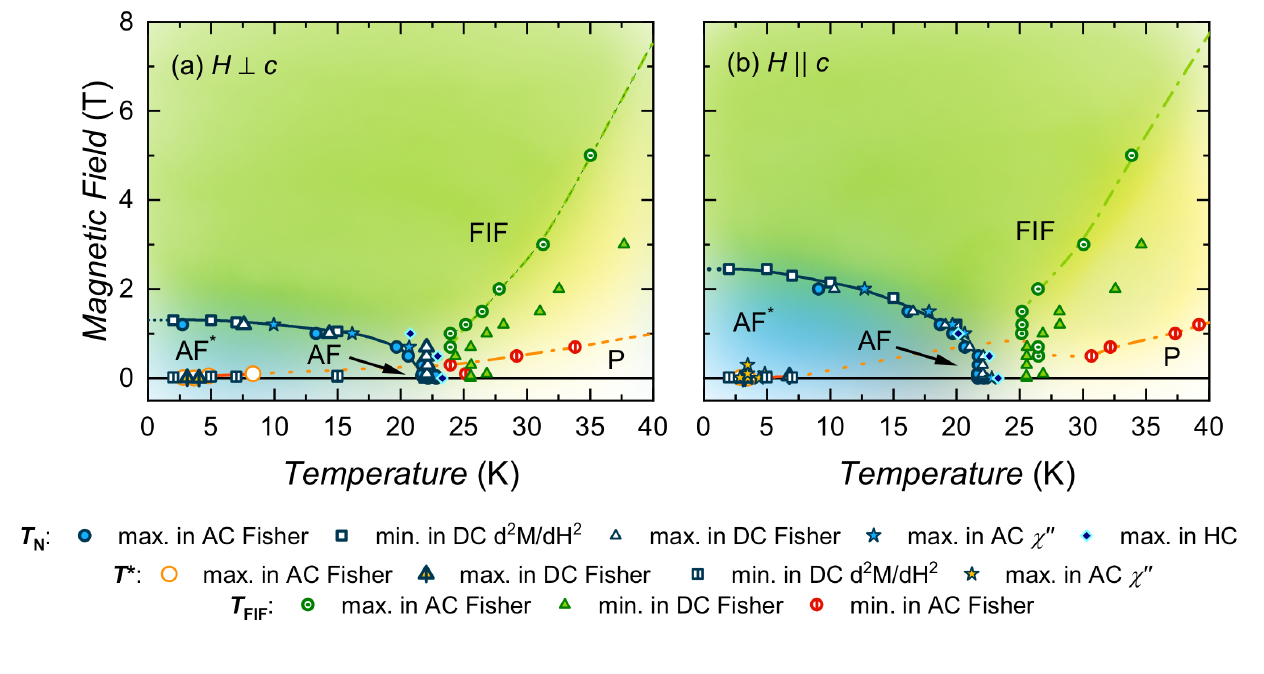
    }
    \caption{Magnetic phased diagram of \EZP{} for magnetic fields applied perpendicular (a) and parallel (b) to the crystallographic $c$ axis. The solid and dashed lines are the guides for the eyes.}
    \label{fig:PhDiag}
\end{figure*}

In Fig. \ref{fig:PhDiag} we summarised characteristic transition points that were obtained as follows.
Using the Fisher's method \cite{Fischer1962}  local maxima and minima in  $d\chi \cdot T/dT$ vs. $T$ [where $\chi$  is either the real part of ac susceptibility $\chi'(T)$ or the dc susceptibility $\chi(T)$ for each magnetic field] were taken as the characteristic transition points (\textsf{max. in AC Fisher} and \textsf{max. in DC Fisher}). The imaginary part of ac susceptibility $\chi''(T)$ for each magnetic field was investigated for local maxima (\textsf{max. in AC $\chi''$}). By taking the minimum in $d^2M/dH^2$ vs. $H$ \textsf{min. in DC $\mathsf{d^2M/dH^2}$} points were determined. The \textsf{max. in HC} points are corresponding to the maxima in heat capacity $C_p(T)$ for each magnetic field.

Based on these data the magnetic phase diagram was constructed, as a result several phase regions can be distinguished for this compound. 
As can be seen in Fig.~\ref{fig:PhDiag} for zero field, going from higher temperatures, a paramagnetic region (\textsf{P}) is followed by a transition at the N\'{e}el temperature  $\TN = \SI{23}{K}$ to antiferromagnetic order with \Euion{} magnetic moments tilted from the basal plane towards the \cax{} (\textsf{AF}). 
At lower temperatures (around $\Tstar=\SI{5}{K}$) a feature shifting to higher temperatures with increasing field,  indicating its origin in ferromagnetic interactions between the \Euion{} moments, is visible (\textsf{AF}$^{*}$ region).
With application of an external magnetic field another region (\textsf{FIF}) can be observed, which is related to the crossover from the paramagnetic state to  field induced alignment of the \Euion{} magnetic moments below the temperature $\TFIF$.

The observed phenomena indicate that the ground state of \EZP{} is an antiferromagnet with moments tilted from the basal plane, however there are some additional ferromagnetic features possibly originating from presence of magnetic domains or canted antiferromagnetic order.

\subsection{Magnetoresistance}

\begin{figure}[!ht]
\centering
\includegraphics[width=0.95\linewidth]{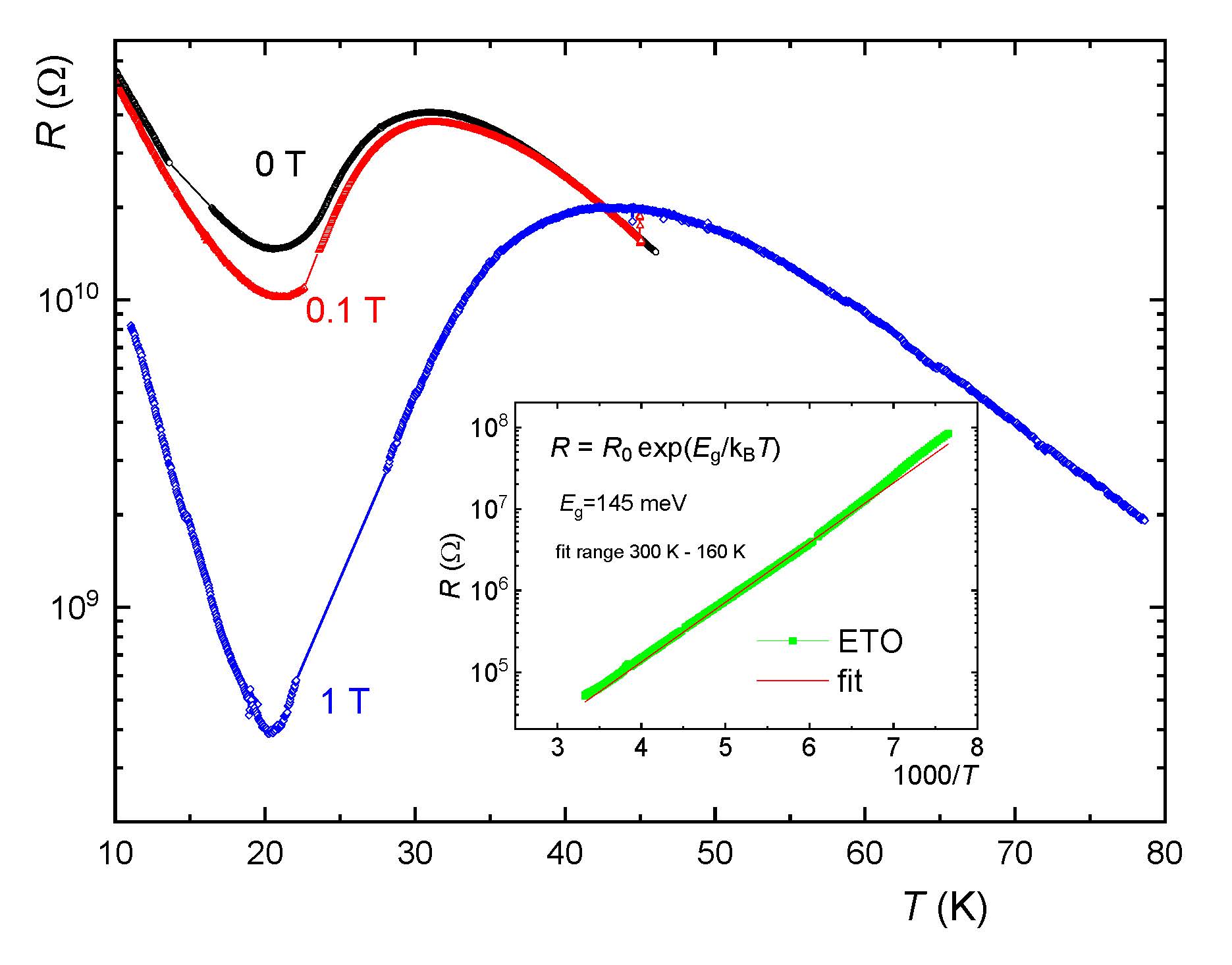}
\caption{Temperature dependence of electrical resistance $R(T)$ for current along the basal plane in various magnetic fields applied parallel to the $c$ axis. The inset shows high temperature data in zero field as a function of $1/T$, used to determine the gap width $E_{g}$. The latter data were obtained using the Electrical Transport Option in PPMS (ETO).}
\label{res1}
\end{figure}

The temperature dependence of electrical resistivity of EuZn$_2$P$_2$ was studied first in Ref.~\citealp{Berry2022}. However, this work covers only a high temperature range (140-\SI{400}{K}). The rapid increase at lower temperatures does not allow to use conventional setups suitable for metallic materials at PPMS systems, as $\si{G\Omega}$ sample resistance is comparable with the resistance of the sample holder without any sample. Here we used an electrometer setup which provides data with reliable temperature dependence as well as absolute values of resistance, which is, however, not easily transformable to resistivity. Very approximately, $\SI{1}{\Omega}$ resistance corresponds for the given geometry to $\rho = \SI{10}{m\Omega.cm}$. On the fine scale, the results differed when changing the range of electrometer, however the orders of magnitude turned out reliable.          
Fig.~\ref{res1} shows that the resistance increases with decreasing $T$ until it reaches a broad maximum related undoubtedly to magnetism, but located still above the actual $\TN$. The $\TN$ actually manifests as a minimum, below which resistance resumes its increasing tendency. The position of the broad maximum is very sensitive to magnetic field, being shifted by more than 10 K (from $T_\text{max} = \SI{30 }{K}$) in magnetic field of \SI{1 }{T}. 
The magnetic phase transition temperature is in agreement with the ac susceptibility measurements, i.e., shifts slowly to lower temperatures with application of an external magnetic field, reaching 20 K in 1 T. 
The related minimum gets deeper and sharper. 

\begin{figure}[!ht]
\centering
\includegraphics[width=0.80\linewidth]{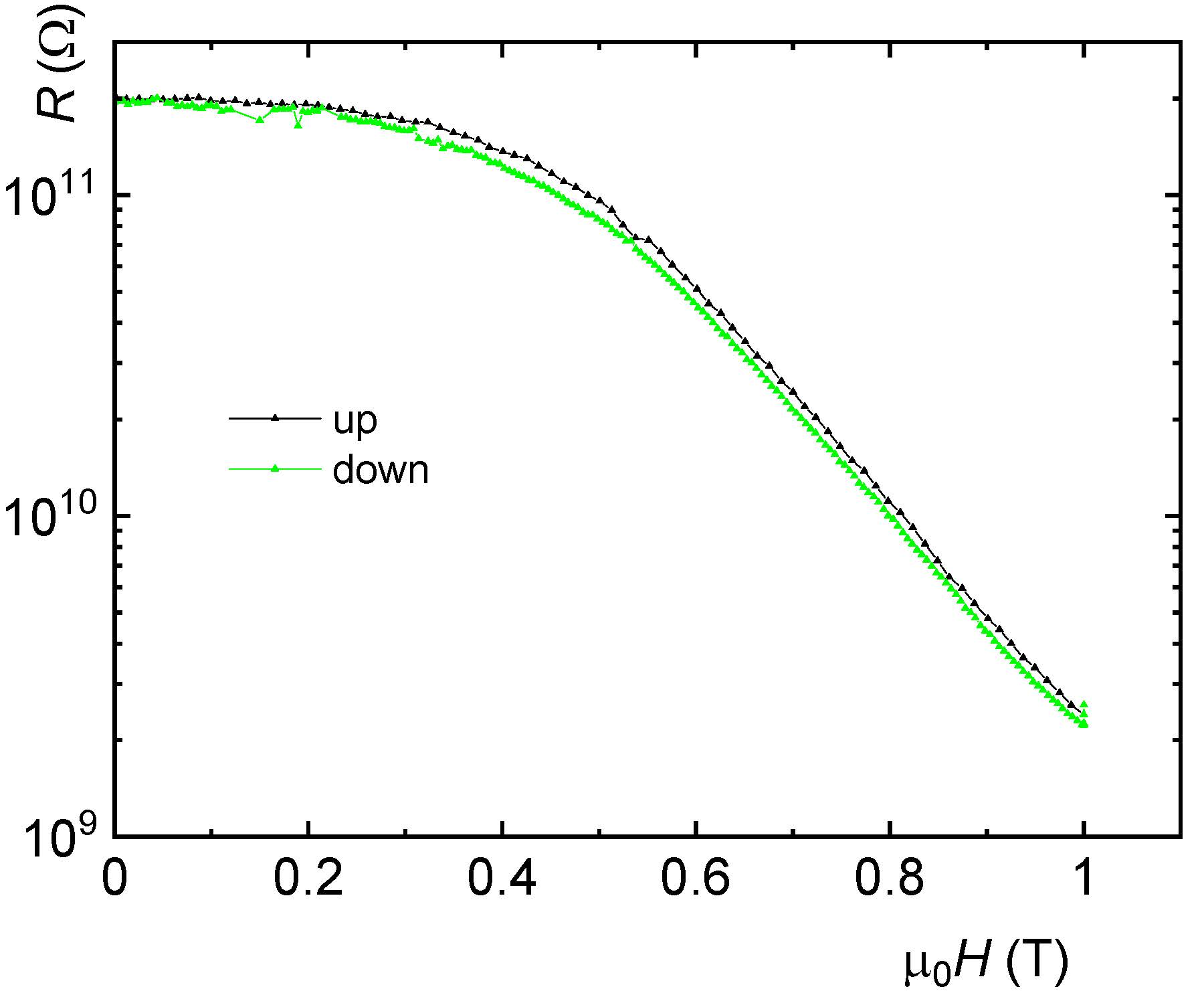}
\caption{Magnetoresistance at $T$ = 10 K of EuZn$_2$P$_2$ for field increasing (black) and decreasing (green). Measurement with current in the basal plane, magnetic field along the \cax{}. }
\label{res2}
\end{figure}

The high-temperature part of $R(T)$ follows an exponential law with the gap width $E_{g} = \SI{0.145}{meV}$, which is not far from the value of \SI{0.11 }{eV} reported in Ref.~\citealp{Berry2022}.
Cooling further below the minimum brings another exponential increase of resistance, and values exceeding $\si{10^{11}.\Omega}$ were recorded at $T = \SI{2}{K}$.
This final increase can be described assuming the gap of merely 2-\SI{3 }{meV}. 
Such situation indicates importance of shallow impurity levels in the gap, affecting the electrical conduction. 
Fig.~\ref{res1} indicates possible colossal magnetoresistance effects, which are indeed disclosed in Fig.~\ref{res2}, revealing a resistance decrease by two orders of magnitude in $\muH = \SI{1}{T}$.

\subsection{High pressure studies}

\begin{figure}[!ht]
\centering
\includegraphics[width=0.90\linewidth]{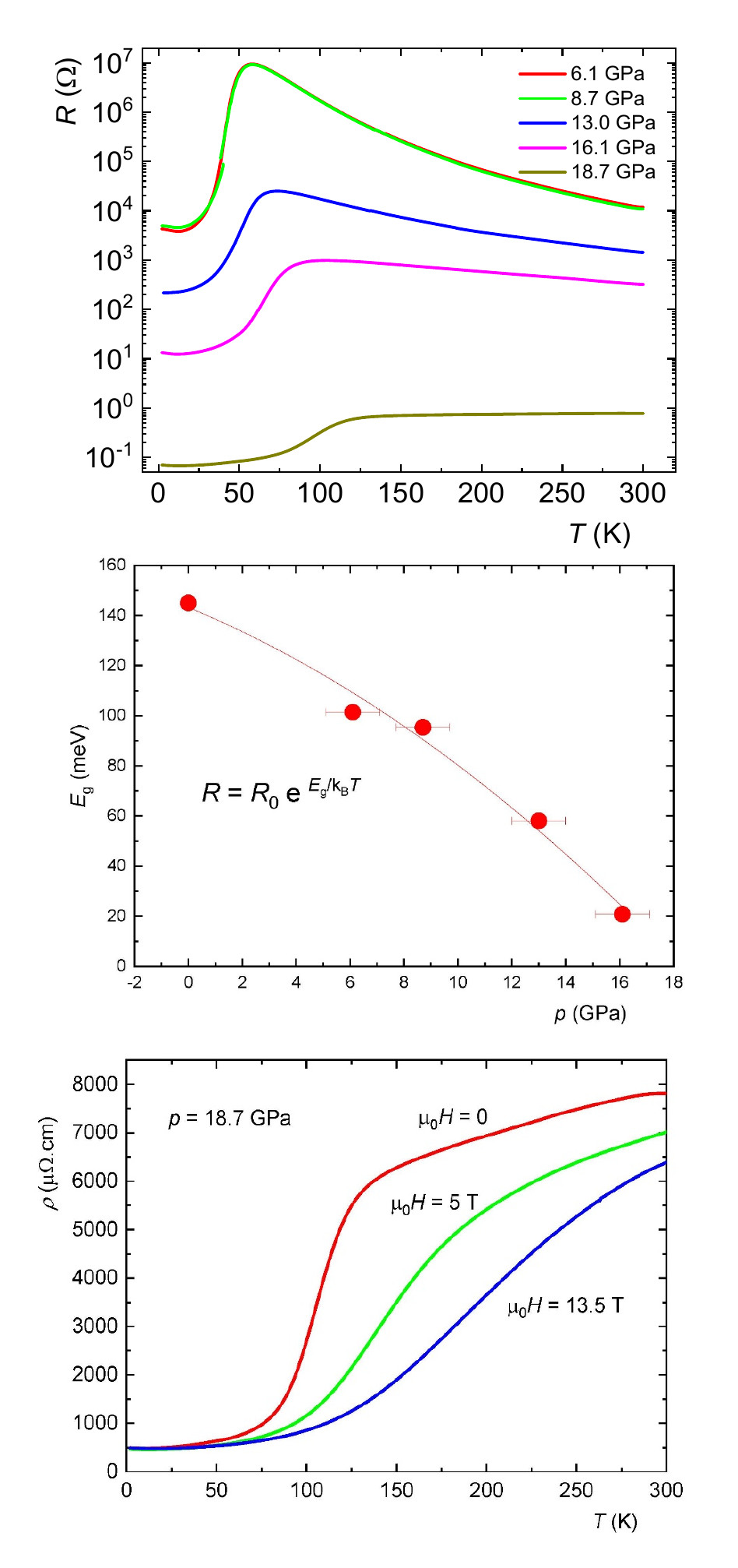}
\caption{ Variations of $R(T)$ of EuZn$_2$P$_2$ as a function of hydrostatic pressure (top). The middle panel describes pressure variations of the gap width. The lower panel displays field variations of resistivity at the highest pressure achieved ($p = \SI{18.7 }{GPa}$), which is sufficient to suppress the gap entirely.}
\label{res3}
\end{figure}

Application of hydrostatic pressure leads to a similar effect as magnetic field. 
The broad maximum of $R(T)$ is gradually shifted towards higher temperatures and absolute $R$-values decrease. 
In addition, the low-$T$ upturn becomes insignificant. 
As seen from Fig.~\ref{res3}, the absolute values decrease by many orders of magnitude. 
The analysis reveals that the activation energy is progressively reduced and the gap disappears at 16-\SI{17 }{GPa}. 
At the highest pressure applied, \SI{18.7 }{GPa}, the sample is apparently in a semi-metallic state already, as indicated by $\rho_{300K} = \SI{8}{m\Omega cm}$ and residual resistivity $\rho_0 = \SI{0.5}{m\Omega cm}$. 
Fig.~\ref{res3} shows that even in such case the resistivity in the paramagnetic state is very strongly field dependent, with large negative magnetoresistance persisting up to the room temperature.
The fact that largest effect is observed around $T = \SI{100}{K}$, where fields of several Tesla affect magnetic fluctuations, indicates the strong coupling of such fluctuations to the pool of conduction electrons. Such effect is necessarily reduced with increasing $T$, as the thermal energy $k_B T$ becomes much higher than the Zeeman energy of a single Eu moment in given fields.       

\begin{figure}[!ht]
\centering
\includegraphics[width=1\linewidth]{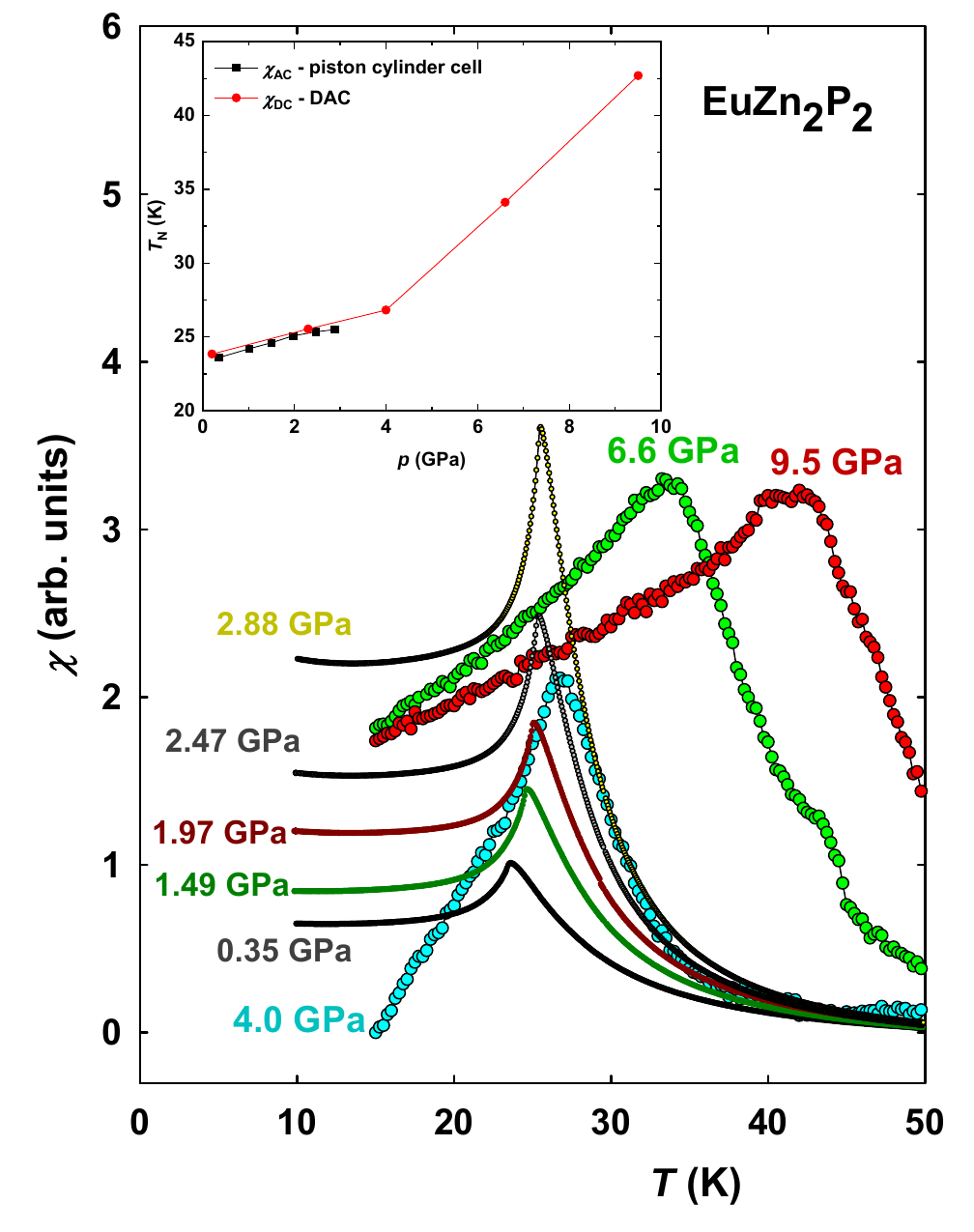}
\caption{ Pressure variations of magnetic susceptibility of EuZn$_2$P$_2$ monitored by \textit{ac} susceptibility in the piston cylinder cell (small points) and dc susceptibility in DAC (large points). The dc susceptibility was measured in the field of \SI{0.01 }{T}. The $\chi(T)$ values were rescaled arbitrarily. The inset displays the pressure dependence of $\TN$ obtained by the two techniques. }
\label{susc_pres}
\end{figure}

As shown above, $\TN$ corresponds at ambient pressure to a minimum in $R(T)$, however, the character of $R(T)$ dramatically changes when pressure is applied and from $R(T)$ it is not obvious what is the pressure variation of $\TN$. 
Therefore, it was essential to determine the susceptibility under pressure. 
In Fig.~\ref{susc_pres} two types of experimental results under pressure are displayed.
The ac susceptibility, $\chi_{ac}(T)$, used in the lower pressure range (up to \SI{3 }{GPa}), provides good hydrostaticity and low noise. The range up to \SI{10 }{GPa} was covered by the dc magnetization measurements. The fact that susceptibility forms a maximum means that the AF order does not transform to ferromagnetism in the pressure range studied. In the range where both techniques overlap we observed their good agreement, following approximately linear increase. This increase becomes steeper around $p = \SI{4 }{GPa}$. The data suggest that there may be another linear increase above this limit, but a more detailed study has to be undertaken to be conclusive. The important fact is that the $\TN$ values at elevated pressures do not correlate with any particular anomaly in $R(T)$. The maximum in $R(T)$ in Fig.~\ref{res3}, albeit related to magnetism, is located above the actual $\TN$ at corresponding pressures. Hence, we tend to associate the maximum, or better the decrease of $R$ from the exponentially increasing tendency, with magnetic fluctuations, inducing magnetic polarons.

As the family of ternary Eu Zintl phases with a distinctly layered structure is quite extended, we can compare the behavior with other systems, having identical or very similar crystal structure. For some of them even the pressure variations of the behavior have been explored. A convenient starting point of understanding is that the Eu magnetism implies the $4f^7$ $5d^0$ (\Euion) state. This charge state is at least to some extent robust against the lattice compression. On the other hand, the state with $4f^6$ $5d^1$ configuration has dramatically lower volume and should be therefore preferred in a highly compressed state. The $4f^6$ state with $L=S=3$, $J = L-S =0$ is naturally non magnetic. In addition, valence fluctuations between one magnetic and one non-magnetic state may also produce non-magnetic ground state due to the dynamics of charge fluctuations or Kondo effect. For systems tunable across the loss of magnetism either by changes of composition or pressure, as e.g. \ce{Eu(Rh_{1-x}Ir_x)2Si2}, the transition can be of the first order type \cite{Seiro2011}. 

A characteristic feature of electronic structure of the Eu Zintl phases (those containing a pnictogen or chalcogen) is existence of a gap at the Fermi level, with the $3p$ states of P contributing to the upper edge of the valence band and $5d$ states of Eu in the conduction band. 

For \EZP, the pressure induced gradual gap suppression, indicated both by calculations and resistivity studies, can thus  eventually bring the $5d$ states to and below the Fermi level. We can speculate then that a crossing of the metal insulator transition giving a fractional occupancy of the $5d$ states may also give a non-magnetic state due to the mean $4f$ occupancy sinking below 7, which would suppress its magnetism. In pressures used here, \ce{EuZn2P2}, is however still on the ascending part of the dependence of $\TN$ on pressure. It implies that the $4f$ occupancy stays at 7 and the effective inter-site exchange coupling increases. As pointed out already for various isostructural compounds \cite{Berry2022}, the Eu-Eu distances are likely the explicit tuning parameter. This is, however, true both for the dipole-dipole interaction suggested in Ref. \cite{Berry2022} and (more likely) for the superexchange, depending on the $4f$-$3p$ hybridization. The observed increase of $\TN$  of \EZP{} under pressure is the continuation of the same trend to even smaller distances than available compounds offer within the same structure type. What remains to complete the story is to describe the pressure variations of lattice parameters. The anisotropy of lattice thermal expansion detected here for \EZP, suggests that also individual compressibilities under hydrostatic pressure may exhibit a significant anisotropy.  

Such robust enhancement of ordering temperatures in Eu systems is naturally not restricted to the given family of ternaries. As a simple model systems one can take, e.g., Eu monochalcogenides, which crystallize in the cubic rocksalt structure. Indeed, magnetic ordering temperatures increase with decreasing size of a chalcogen or with pressure. Moreover, the gradual compression turns the antiferromagnetic order of EuTe into ferromagnet, with the critical Eu spacing around $\SI{5.2}{\angstrom}$, achieved in approx. \SI{8}{GPa} \cite{Ishizuka1997}. On the other side of the series EuO with the ferromagnetic ground state 	has $\TC$ dramatically increasing by \SI{5}{K/GPa} \cite{Stevenson1965} from \SI{69.2}{K} at ambient pressure. A remarkable fact is that both the predominance of the F order at low Eu-Eu distances and increase of $\TC$ with compression are well accounted by \textit{ab initio} calculations (LDA+\textit{U}). \cite{Kunes2005} An analysis shows an important role of the $p$-$f$ mixing and superexchange. With such background information it is understandable that the exchange in the basal plane is strong and ferromagnetic, as the Eu-Eu distances are far below \SI{5}{\angstrom}, while much longer bonds along the \cax{} bring AF coupling, which is so weak that it can be broken in magnetic fields of few Tesla. The superexchange operates, as pointed out in Ref. \cite{Singh2023}, via the P atoms adjacent to the Eu planes. The situation with strong in-plane coupling and weak inter-plane one brings 2D magnetic correlations existing above $\TN$. Their footprint can be seen in magnetic entropy extending up to $T = \SI{40}{K}$ even in zero magnetic field.  

Among compounds with the same crystal structure, \ce{EuZn2As2} with lattice parameters somewhat higher and $\TN = \SI{19.2}{K}$ lower than for \EZP, exhibits a remarkable similarity to \EZP \cite{Bukowski2022}. While there are no high pressure data for \ce{EuZn2As2}, high pressure effects were studied for \ce{EuIn2As2} with a similar structure type~\cite{Yu2020}. It is more metallic than \ce{EuZn2P2}, possibly with a very narrow gap around the $\Gamma$ point, at which nontrivial topologically protected states are present at the Fermi energy $\EF$. As those states depend on the symmetry imposed by magnetic ordering, any moments reorientation yields a large impact on electrical resistivity. A complicated helicoidal magnetic structure of the ground state makes the situation quite involved \cite{Riberolles2021}. The impact of pressure is very dramatic. The $\TN$ value increases up to $\SI{65}{K}$ in $p = \SI{14.7}{GPa}$, which is attributed to stronger exchange coupling in the basal plane. In pressures exceeding \SI{17}{GPa} it becomes amorphous and metallic, so the trend could not be followed continuously any more \cite{Yu2020}.  We can think about the first order transition coupling the volume collapse with the valence change. The amorphization can appear in the situation in which a local structure motif imposed by bonding conditions modified by the compression is incompatible with the 3D translation periodicity of the crystal lattice. An interesting feature is rather low bulk modulus $B_0 = \SI{32.3}{GPa}$. Comparing the linear compressibilities from the data in  Ref. \cite{Yu2020} one can deduce that the \cax{} is actually much softer.  Increasing pressure brings metallization seen in electrical resistivity, while the colossal magnetoresistance effect is gradually suppressed. 
Although topological states depending on the symmetry of the AF state turning to F under the influence of field can affect the resistivity, as general and more simple mechanism is a reduction of the band gap. Even if it happens for one spin direction only, a simple parallel resistor model reveals that the opposite spin direction where the gap increases becomes for conduction practically irrelevant. The other side of the coin is that the conduction electrons are very strongly spin polarized, which gives rise to a spin filter effect, described, e.g., for EuO \cite{Jutong2012}. An interesting fact is that the largest magnetoresistance effect is for \ce{EuZn2P2} recorded not in the low temperature limit, but above $\TN$, where the resistance forms a pronounced knee. Evidently the strong coupling of transport to magnetic fields reflects the field effect on magnetic correlations above $\TN$. In particular, we can expect magnetic field to have a stabilizing effect on ferromagnetic fluctuations, assumed within the basal plane. As we measure, due to the geometry of the crystals grown, the resitivity perpendicular to the \cax, the experiment is particularly sensitive to the correlations in the basal plane. Applying pressure the knee shifts accordingly, exceeding $T = \SI{100}{K}$. The largest magnetoresistance point shifts to approx. \SI{150}{K}. The fact that there is very little field dependence in the low-$T$ limit can mean that the type of order actually changes from AF to F in the higher pressures applied in this work.

\section{Summary and conclusions}
\ce{EuZn2P2} is an antiferromagnetic layered compound with a narrow band gap, which is intimately connected with the magnetic state.  We assume a strong ferromagnetic Eu-Eu exchange coupling  within the basal plane sheets, based on superexchange including phosphorus ions. The weaker interaction to more remote Eu neighbours along the \cax{} is antiferromagnetic, which determines the type of magnetic order.
The \Euion{} moments are inclined from the basal plane.  Large Eu moments of $\SI{7}{\mu_B}$, 
weak AF coupling and very weak magnetocrystalline anisotropy (\Euion{} has spin-only moments similar to \ce{Gd^{3+}}, but absence of $5d$ occupancy makes the anisotropy even weaker than for Gd systems) facilitate the rotation of moments into a field aligned state, accomplished in $\muH = \SI2{T}$. This dramatically reduces resistivity not only in the AF state but also above $\TN$. Such CMR effect is even more pronounced in a limited $T$ range above the actual $\TN$, which is related to ferromagnetic correlations. The phenomenon can be related to the reduction of the gap width. The gap is suppressed in $p \approx \SI{17}{GPa}$, where \ce{EuZn2P2} becomes a semi-metal, probably ferromagnetic. Hence we can anticipate that it is also a half-metal with one spin direction only at the Fermi level. Hydrostatic pressure also increases $\TN$ from \SI{23}{K} to almost \SI{45}{K} in $p = \SI{10}{GPa}$, which seems to be quite general effect for Eu systems with a band gap \cite{Stevenson1965}. 
As the crystal structure does not have a center of symmetry, further research will have to focus on possible non-trivial topological states and their interplay with the symmetry of magnetic state and its development in magnetic fields.


\section*{Acknowledgements}

D. R. acknowledges financial support by National Science Centre, Poland (Grant No. 2018/30/E/ST3/00377). W. T. and K. K. acknowledge the support of the National Science Centre, Poland, Grant No. OPUS: 2021/41/B/ST3/03454.
Research project was partly supported by program „Excellence initiative – research university” for the AGH University of Krakow.
 The work has been supported by the Czech Science Foundation under the grants No. 21-09766S and 22-22322S (DL).  DL also acknowledges the computational resources by project e-INFRA CZ (ID:90254) and DL project QM4ST CZ.02.01.01/00/22\_008/0004572 both by the Ministry of Education, Youth and Sports of the Czech Republic. Some experiments were performed in MGML (mgml.eu), which is supported within the program of the Czech Research Infrastructures (project no. LM2023065).






\bibliography{a_EuZn2P2}

\end{document}


\title{Supplementary Material: Ambient and high pressure studies of structural, electronic and magnetic properties of EuZn$_2$P$_2$ single crystal}

\author{Damian Rybicki}
\affiliation{AGH University of Krakow, Faculty of Physics and Applied Computer Science, al. A. Mickiewicza 30, 30-059 Krak\'ow, Poland}

\author{Kamila Komędera}
\affiliation{AGH University of Krakow, Faculty of Physics and Applied Computer Science, al. A. Mickiewicza 30, 30-059 Krak\'ow, Poland}

\author{Janusz Przewoźnik}
\affiliation{AGH University of Krakow, Faculty of Physics and Applied Computer Science, al. A. Mickiewicza 30, 30-059 Krak\'ow, Poland}

\author{Łukasz Gondek}
\affiliation{AGH University of Krakow, Faculty of Physics and Applied Computer Science, al. A. Mickiewicza 30, 30-059 Krak\'ow, Poland}

\author{Czesław Kapusta}
\affiliation{AGH University of Krakow, Faculty of Physics and Applied Computer Science, al. A. Mickiewicza 30, 30-059 Krak\'ow, Poland}

\author{Karolina Podgórska}
\affiliation{AGH University of Krakow, Faculty of Physics and Applied Computer Science, al. A. Mickiewicza 30, 30-059 Krak\'ow, Poland}

\author{Wojciech Tabiś}
\affiliation{AGH University of Krakow, Faculty of Physics and Applied Computer Science, al. A. Mickiewicza 30, 30-059 Krak\'ow, Poland}

\author{Jan Żukrowski}
\affiliation{AGH University of Krakow, Faculty of Physics and Applied Computer Science, al. A. Mickiewicza 30, 30-059 Krak\'ow, Poland}

\author{Lan Maria Tran}
\affiliation{Institute of Low Temperature and Structure Research, Polish Academy of Sciences, Wroc\l{}aw, 50-422, Poland}
\author{Micha\l{} Babij}
\affiliation{Institute of Low Temperature and Structure Research, Polish Academy of Sciences, Wroc\l{}aw, 50-422, Poland}
\author{Zbigniew Bukowski}
\affiliation{Institute of Low Temperature and Structure Research, Polish Academy of Sciences, Wroc\l{}aw, 50-422, Poland}

\author{Ladislav Havela}
\affiliation{Department of Condensed Matter Physics, Faculty of Mathematics and Physics, Charles University, Ke
Karlovu 3, 121 16 Prague 2, Czech Republic}

\author{Volodymyr Buturlim}
\affiliation{Department of Condensed Matter Physics, Faculty of Mathematics and Physics, Charles University, Ke
Karlovu 3, 121 16 Prague 2, Czech Republic}

\author{Jiri Prchal}
\affiliation{Department of Condensed Matter Physics, Faculty of Mathematics and Physics, Charles University, Ke
Karlovu 3, 121 16 Prague 2, Czech Republic}

\author{Martin Divis}
\affiliation{Department of Condensed Matter Physics, Faculty of Mathematics and Physics, Charles University, Ke
Karlovu 3, 121 16 Prague 2, Czech Republic}

\author{Petr Kral}
\affiliation{Department of Condensed Matter Physics, Faculty of Mathematics and Physics, Charles University, Ke
Karlovu 3, 121 16 Prague 2, Czech Republic}

\author{Ilja Turek}
\affiliation{Department of Condensed Matter Physics, Faculty of Mathematics and Physics, Charles University, Ke
Karlovu 3, 121 16 Prague 2, Czech Republic}

\author{Itzhak Halevy}
\affiliation{Ben Gurion university, Be’er Sheva, Israel}

\author{Jiri Kastil}
\affiliation{Institute of Physics, Academy of Science of the Czech Republic, Prague 8, Czech Republic}

\author{Martin Misek}
\affiliation{Institute of Physics, Academy of Science of the Czech Republic, Prague 8, Czech Republic}

\author{Urszula D. Wdowik}
\affiliation{IT4Innovations, V\v{S}B - Technical University of Ostrava, 17. listopadu 2172/15, 70800 Ostrava, Czech Republic} 

\author{Dominik Legut}
\affiliation{Department of Condensed Matter Physics, Faculty of Mathematics and Physics, Charles University, Ke
Karlovu 3, 121 16 Prague 2, Czech Republic}
\affiliation{IT4Innovations, V\v{S}B - Technical University of Ostrava, 17. listopadu 2172/15, 70800 Ostrava, Czech Republic}

\maketitle

\section{Crystals}

\begin{figure}[h]
    \centering
    \includegraphics[width=0.5\textwidth]{Rys/SM_Crystals.jpg}
    \caption{Typical single crystals of \EZP{} on a millimeter grid.}
    \label{fig:Crystals}
\end{figure}

\begin{figure}[h]
    \centering
    \includegraphics[width=1\textwidth]{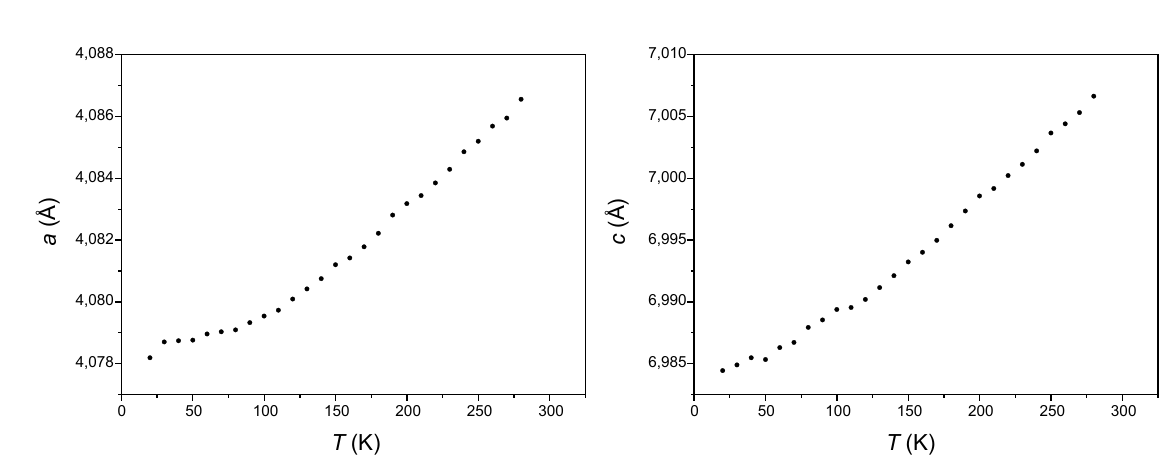}
    \caption{Temperature dependence of the lattice parameters $a$ and $c$.}
    \label{XRD_suppl}
\end{figure}

Fig.~\ref{fig:Crystals} exhibits the size and morphology of some of the single crystals of \EZP{}. 
Fig. \ref{XRD_suppl} presents the temperature dependence of the lattice parameters $a$ and $c$ as determined by XRD. 

\section{Ab-initio calculations}

\begin{figure}[h]
    \centering\includegraphics[width=0.60\linewidth]{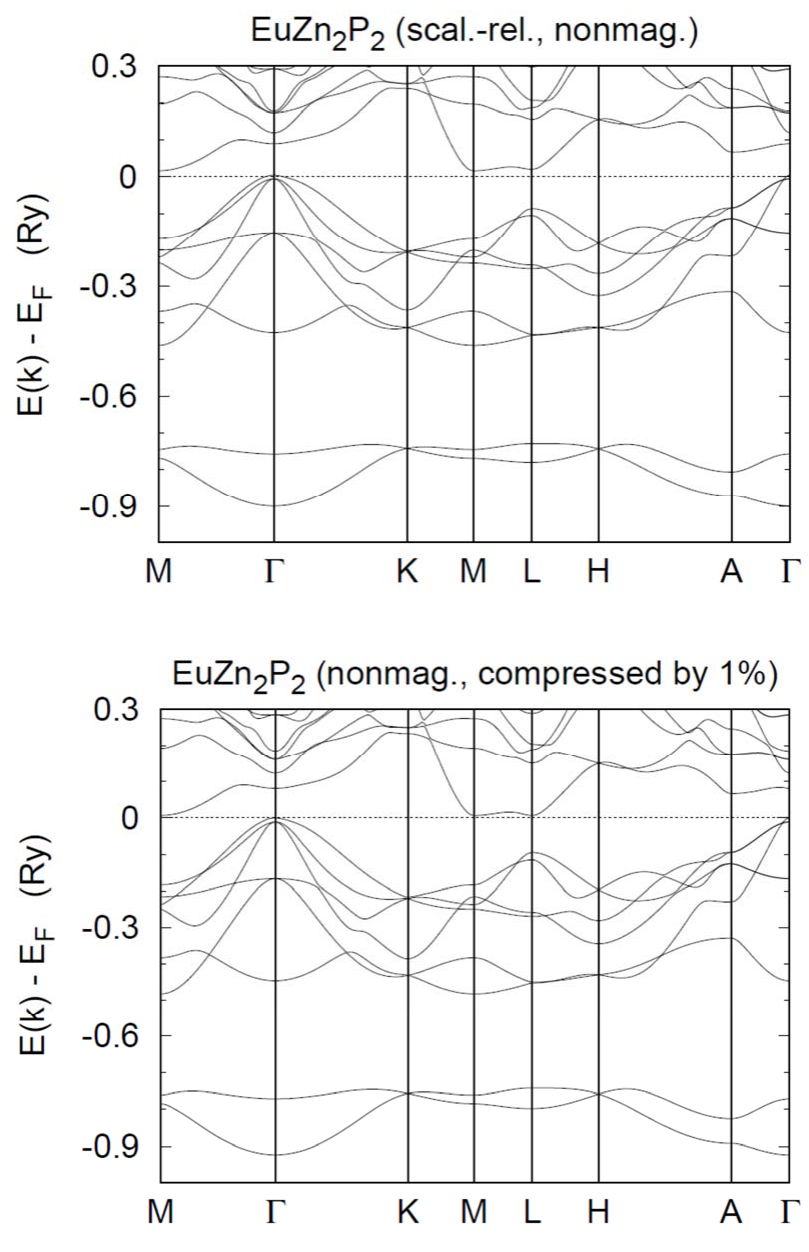}
    \caption{Electronic structure of EuZn$_2$P$_2$ obtained using the scalar relativistic LMTO method (LDA) for the equilibrium volume (top) and 3\% volume (1\% linear) compression  (bottom).}
    \label{Turek_compress}
\end{figure}

Several different ab-initio methods were used to characterize the electronic structure and lattice dynamics. One of important features is the gap in the electronic spectrum. Practically all the methods employed indeed yield a gap, although its width varied depending on the method used. Our aim was first to identify whether there is a general tendency for the gap suppression with a volume reduction, produced by hydrostatic pressure. Hence we used first a simple non-magnetic scalar relativistic tight-binding LMTO method \cite{turek_electronic_1997}\cite{pajda_oscillatory_2000}\cite{turek_exchange_2006} with the 4\textit{f} states treated as open core. This approach, which suppresses the hybridization with non-\textit{f} states and fixes the 4\textit{f} occupancy to 7, is comparable with practically full localization of the 4\textit{f} states.  Within this framework we compared the equilibrium state with experimental lattice parameters with the state isotropically compressed by to 1\% in each direction, corresponding to 3\% volume compression. The  ambient pressure situation yields a gap of 164 meV, which is reduced to 88 meV with volume compression of 3\%. The dispersion relations shown in Fig. \ref{Turek_compress} reveal that such numbers are attributed to the indirect gap. The direct gap at the $\Gamma$ point is an order of magnitude higher. Our results offer qualitatively similar picture to results of Ref. \cite{Berry2022} with lowest unoccupied states at the M and L points of the Brillouin zone, which are manifest in the indirect gap. Extending the calculation to LSDA+\textit{U}, in which \textit{U} = 6 eV provides the necessary 4\textit{f} localization (not shown here), gives a ferromagnetic state with the indirect gap $\Delta$ = 283 meV (between $\Gamma$ spin-down and L spin-up states), which is again reduced to 203 meV upon 3\% volume compression. 

\begin{figure}[h]
    \centering
\includegraphics[width = 0.9\linewidth]{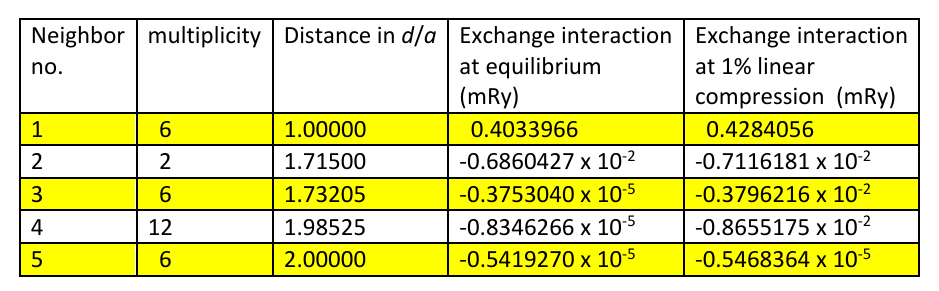}
    \caption{Summary of individual pair exchange interaction parameters of EuZn$_2$P$_2$ in equilibrium volume and at 3\% volume (1\% linear compression). The distances within each basal-plane sheet marked in yellow, other distances are between Eu atoms in adjacent sheets.}
    \label{exchange_table_cropped}
\end{figure}

Within this approach we also performed mapping to an isotropic Heisenberg model with pair exchange interactions between the Eu moments. Their  evaluation employed the magnetic force theorem \cite{liechtenstein_local_1987}  implemented within the scalar-relativistic TB-LMTO method. 
The resulting pair interactions for Eu-Eu distances not exceeding 2\textit{a} (up to the 5th nearest neighbors) are tabulated in Fig. \ref{exchange_table_cropped}. One can see the dominating ferromagnetic 1st nearest-neighbor interaction \textit{J}1, which is two or more orders of magnitude stronger than interactions between more distant neighbors, which are antiferromagnetic. Note that the interactions \textit{J}1, \textit{J}3, and \textit{J}5 refer to pairs lying in the same basal plane while \textit{J}2 and \textit{J}4 couple atomic pairs located in two neighboring basal planes. These results prove that the system behaves essentially as a collection of separated two-dimensional ferromagnets with nearest-neighbor coupling \textit{J}1. Upon the isotropic volume compression all the pair exchange parameters increase in absolute value.

A more complete study of lattice compression was performed using the fully relativistic FPLO  method \cite{eschrig_density_2003} (based on 4-component Dirac equation) using the  GGA + \textit{U} approximation, with \textit{U} taken also as 6 eV, Fully Localized Limit - FLL treatment of double counting, assuming a ferromagnetic state with Eu moments along the \textit{c}-axis.  Although in this case the gap comes out  wider ($\Delta$ = 470 meV) than in the previous case, it is gradually suppressed by hydrostatic pressure, and the estimated critical pressure for insulator-metal transition is 12 GPa. The calculations give the equilibrium volume by 1.0\% higher than the experimental one and the bulk modulus \textit{B} = 67 GPa. In these calculations the internal parameters of Zn and P atoms were fixed to values obtained from the force minimization for the experimental equilibrium. 

The dynamics of crystal lattice was explored by means of the calculations using the 
the Vienna Ab initio Simulation Package (VASP) \cite{VASP}
with the projector augmented wave scheme.\cite{Blochl1994,Kresse1999}
Electron exchange and correlation potentials were treated
within the generalized gradient approximation of Perdew-Burke-Ernzerhof (GGA-PBE). \cite{PBE} Valence electrons for 
Europium atom were in the $6s^2$ $5s^2$ $5p^6$ $4f^7$ shells, in the shells $4p^2$ $3d^{10}$ for Zn, and in the $3s^2$ $3p^3$
configuration for phosphorus.

\begin{figure}[h]
    \centering\includegraphics[width=0.45\linewidth]{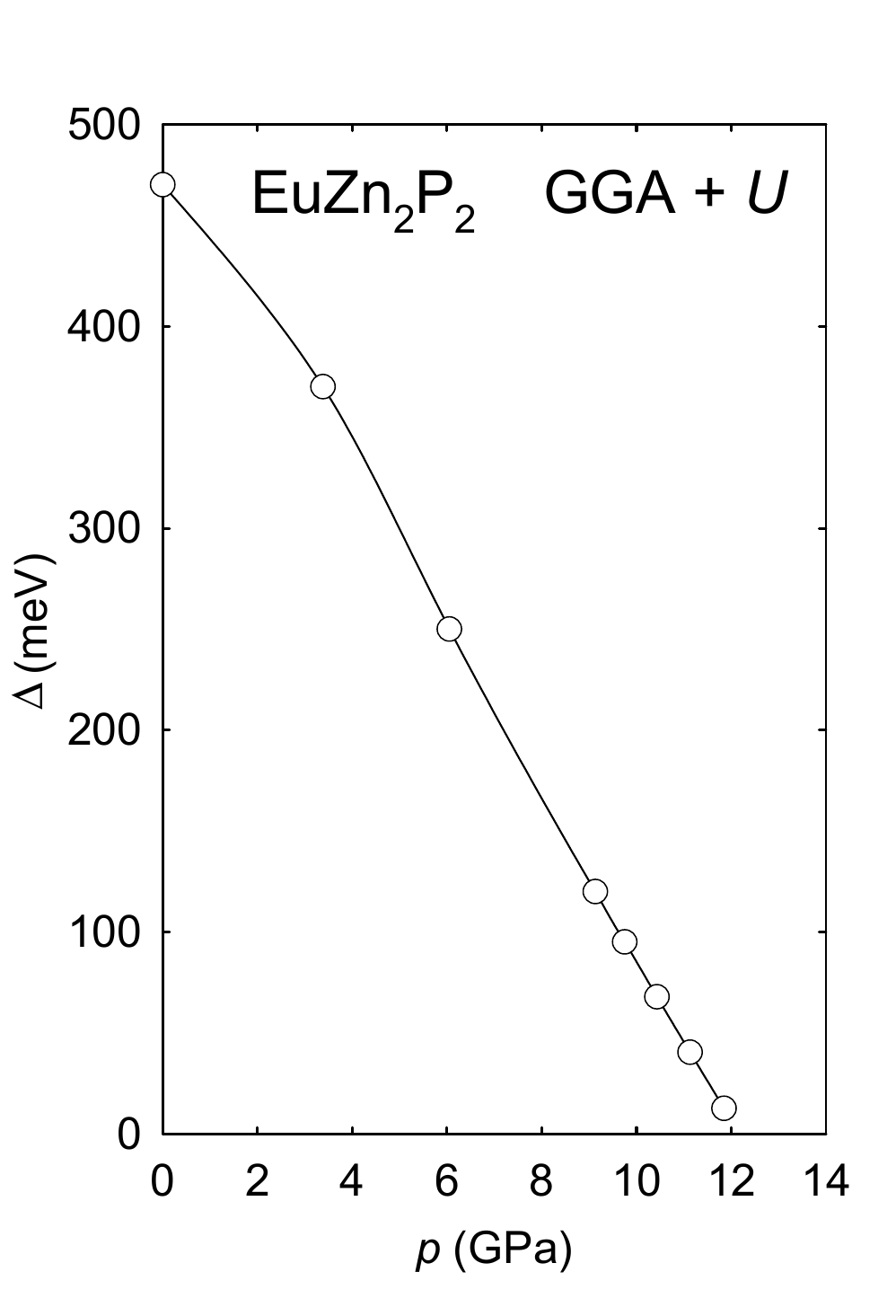}
    \caption{Width of the gap of  EuZn$_2$P$_2$ calculated using the FPLO method within the GGA + \textit{U} approximation. Fee full text for more details.}
    \label{gap_pressure}
\end{figure}

\begin{figure}[h]
    \centering
    \includegraphics[width = 0.6\linewidth]{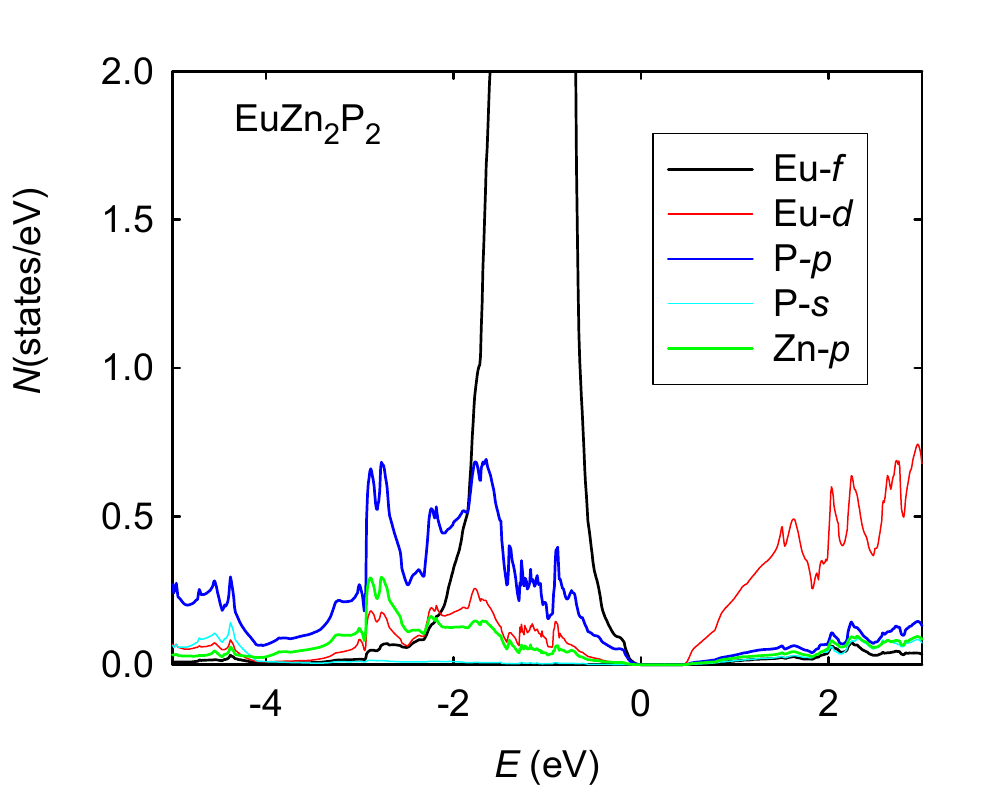}
    \caption{Selected partial densities of electronic states (DOS) of EuZn$_2$P$_2$ obtained by the GGA+\textit{U} calculations. The most important types of states are marked by bold lines. One can see the gap approx. 400 meV wide. The states below the gap are dominated by the Eu-4\textit{f} and P-3\textit{p} states, while the Eu-5\textit{d} states dominate above the gap. The Fermi energy is located at \textit{E} = 0. The 4\textit{f} DOS exceeds significantly the frame of the figure.}
    \label{Legut_DOS}
\end{figure}

The kinetic energy cutoff for the plane-wave basis functions is 750 eV. Brillouin zone integrations were carried out
using a $12 \times 12 \times 6$-centered k-point mesh. We include the spin-orbit coupling (SOC) and apply
the Hubbard model in the rotationally invariant approach \cite{Liechtenstein1995}
with $J = \SI{1.0}{eV}$ and Hubbard $U = \SI{7.5}{eV}$ based in order to get reproduce correctly the lattice parameters as well as the second derivative of total energy vs. volume (bulk modulus). Assuming for simplicity a ferromagnetic ordering, we obtain magnetic moments $\mu = 6.8 \muB$/Eu, $0.1 \muB$/Eu of which is related to a spin polarization of Zn and P. The converged projections of Eu moments are equal in all three Cartesian coordinates. The obtained DOS is displayed in Fig. \ref{Legut_DOS}. In this case the gap width is approx. 400 meV. The given method treats the 4\textit{f} states as narrow-band states, with the band width affected, besides the spin-orbit interaction also by hybridization with non-\textit{f} states. Therefore this technique is not capable to describe correctly the many-body phenomena at the verge between localization and valence fluctuation regime. The small occupancy of the Eu-5\textit{d} states (seen in Fig. \ref{Legut_DOS}, which are expected to be empty in the magnetic 4\textit{f}$^7$, 5\textit{d}$^0$ Eu$^{2+}$ 
state can be taken actually as an artifact, correlating with the 4\textit{f} moments come out slightly smaller than 7 $\muB$.

\clearpage
\section{Magnetic measurements}
\subsection{AC susceptibility}

For easier comparison of the values between measurements in each field, temperature dependencies of AC susceptibility measured in external magnetic fields up to \SI{0.2}{T} are presented in Fig. \ref{fig:AC-compare}. 
\begin{figure}[!ht]
    \centering
    \includegraphics[width = 0.9\linewidth]{Rys/SM_AC-compare.pdf}
    \caption{Temperature dependence of AC magnetic susceptibility investigated in external magnetic fields applied perpendicular (a) and parallel (b) to the crystallographic $c$  axis.}
    \label{fig:AC-compare}
\end{figure}

Comparison between temperature dependence of ac susceptibilities of \ce{EuFe2As2} and \ce{EuZn2P2} measured in magnetic field applied perpendicular and parallel to the crystallographic \cax{} is presented in~Fig.~\ref{fig:AC-Eu-122-Ha} and Fig.~\ref{fig:AC-Eu-122-Hc}, respectively. While the $\chi'(T)$ and $\chi''(T)$ measured for $H\perp c$ have similar features (Fig.Z\ref{fig:AC-Eu-122-Ha}), there is no resemblance for the measurements in $H \parallel c$ (Fig.~\ref{fig:AC-Eu-122-Hc}).

\begin{figure}[!ht]
    \centering
    \includegraphics[width=0.49\linewidth]{Rys/SM_AC_EuFe2As2_Ha.pdf}
    \includegraphics[width=0.49\linewidth]{Rys/SM_AC_EuZn2P2_Ha.pdf}    
    \caption{Temperature dependence of real $\chi'(T)$ and imaginary $\chi''(T)$ parts of AC susceptibility of \ce{EuFe2As2} (a-d) and \EZP{} (e-h) investigated in external magnetic field $\muH$ applied perpendicular to the crystallographic \cax.}
    \label{fig:AC-Eu-122-Ha}
\end{figure}

\begin{figure}[!ht]
    \centering
    \includegraphics[width=0.49\linewidth]{Rys/SM_AC_EuFe2As2_Hc.pdf}
    \includegraphics[width=0.49\linewidth]{Rys/SM_AC_EuZn2P2_Hc.pdf}
    
    \caption{Temperature dependence of real $\chi'(T)$ and imaginary $\chi''(T)$ parts of AC susceptibility of \ce{EuFe2As2} (a-d) and \EZP{} (e-h) investigated in external magnetic field $\muH$ applied parallel to the crystallographic \cax.}
    \label{fig:AC-Eu-122-Hc}
\end{figure}

\clearpage

\subsection{DC magnetization}

Temperature dependence of DC magnetic susceptibility investigated for various fields is presented in Fig.~\ref{fig:DCSusc-compare}  and  Fig.~\ref{fig:DCSusc-total}. 

In~Fig. \ref{fig:DCSusc-compare} only dependencies for up to \SI{0.1}{T} are presented. In~Fig.~\ref{fig:DCSusc-total} susceptibilities for fields up to \SI{9}{T} are presented, however for better visibility the subsequent dependencies are shifted along the $y$ axis by 
\SI{-e-5}{m^3 mol_{at. Eu}^{-1}}.
\begin{figure}[!htb]
    \centering
    \includegraphics[width=0.9\linewidth]{Rys/SM_DCSusc-compare.pdf}
    \caption{Temperature dependence of DC magnetic susceptibility investigated in external magnetic fields applied perpendicular (a) and parallel (b) to the $c$  axis.}
    \label{fig:DCSusc-compare}
    \end{figure}
    
    \begin{figure}[!htb]
    \centering
    \includegraphics[width = 0.9\linewidth]{Rys/SM_DCSusc-total.pdf}
    \caption{Temperature dependence of DC magnetic susceptibility investigated in external magnetic fields applied perpendicular (a) and parallel (b) to the $c$  axis. The dependencies for $\muH\geq \SI{0.01}{T}$ are shifted along $y$-axis by \SI{-e-5}{m^3 mol_{at. Eu}^{-1}}.}
    \label{fig:DCSusc-total}
\end{figure}

\clearpage

\bibliography{a_SuppMat}